\documentclass{ws-ijbc}
\usepackage[english]{babel}
\usepackage{graphicx}
\begin{document}

\catchline{}{}{}{}{} 

\markboth{Barrio, Shilnikov and Shilnikov}{Kneadings, symbolic dynamics and painting Lorenz chaos}

\title{KNEADINGS, SYMBOLIC DYNAMICS AND PAINTING LORENZ CHAOS. A TUTORIAL}

\author{ROBERTO BARRIO}
\address{Departamento de Matem\'atica Aplicada and IUMA,
University of Zaragoza. E-50009. Spain. \\ Email: \texttt{rbarrio@unizar.es.}}

\author{ANDREY SHILNIKOV}
\address{Neuroscience Institute and Department of  Mathematics
and Statistics, Georgia State University, Atlanta 30303, USA. Email: \texttt{ashilnikov@gsu.edu.}}

\author{LEONID SHILNIKOV}
\address{Institute for Applied Mathematics and Cybernetics,
Nizhny Novgorod, 603005 Russia.}

\maketitle

\begin{history}
\received{(to be inserted by publisher)}
\end{history}

\begin{abstract}
A new computational technique based on the symbolic description utilizing kneading invariants
is proposed and verified for explorations of dynamical and parametric chaos in a few exemplary systems with the Lorenz attractor.
The technique allows for uncovering the stunning complexity and universality of bi-parametric structures and detect their organizing centers -- codimension-two T-points and separating saddles in the kneading-based scans of the iconic Lorenz equation from hydrodynamics, a normal model from mathematics, and a laser model from nonlinear optics.
\end{abstract}



\section{Introduction}\label{sec:1}
A great deal of analytical and experimental studies, including modeling simulations, have been focused
on the identification of  key signatures to serve as structural invariants that would allow dynamically alike nonlinear systems
with chaotic dynamics from diverse origins to be united into a single class. Among these key structures are various homoclinic and heteroclinic
bifurcations of low codimensions that lie at the heart of the understanding of complex behaviors because of their roles of
organizing centers of dynamics in parameterized dynamical systems.

Dynamical systems theory has aimed to create purely abstract approaches that are further proceeded by creation and
the development of applicable tools designed for the search and identification of such basic invariants for simple, Morse-Smale, systems and ones with  complex chaotic dynamics.
One such a [computationally justified] approach for studying complex dynamics capitalizes on the concept of sensitivity of deterministic chaos.
Sensitivity of chaotic trajectories can be quantified in terms of the divergence rate evaluated through the largest Lyapunov characteristic exponent. The approach has been proven to work exceptionally well for various systems with chaotic and simple dynamics.
In several low-order dissipative systems, like the R\"ossler model, the computational technique based on  the largest Lyapunov characteristic exponent
reveals that they possess common, easily recognizable patterns involving spiral structures in bi-parametric planes
\cite{ABS77,BYK93,ALS91,SST93,BBSS11,Gal10}. Such patterns turn out to be ubiquitously alike in both time-discrete \cite{Lor08} and time-continuous systems \cite{GN84,BBS09a,Gal10}, and they are easily located when the spiral patterns have regular and chaotic spiral ``arms'' in the systems with the Shilnikov saddle-focus \cite{LPALS07}.

Application of the Lyapunov exponents technique fails, in general, to reveal fine structures embedded in the bi-parametric scans of
Lorenz-like systems. As such it cannot deliver the desired insights into intrinsic bifurcations because regions of chaotic dynamics appear
to be uniformly.  This basically means that the instability of the Lorenz attractors does not vary noticeably as control parameters of the system are varied. This holds true too when one attempts to find the presence  of characteristic spiral structures that are known to exist theoretically in the Lorenz-like systems \cite{BYK93,GS86} and therefore could only be identified using accurate bifurcation continuation approaches \cite{ALS91,SST93}.
Such spirals in a bi-parametric parameter plane of the system in question
are organized around the so-called  T[erminal]-points corresponding to codimension-two or -higher heteroclinic connections between two or more saddle equilibria.
For $\mathbb{Z}_2$-symmetric systems with the Lorenz attractor, the degeneracy is reduced to a cod-2 bifurcation of a closed heteroclinic connection involving two saddle-foci and a saddle at the origin. T-points have been located in various models of diverse origins including electronic oscillators \cite{Bykov98,FFR02} and nonlinear optics \cite{FMH91,WK05}, etc.

\begin{figure}[hbt!]
 \begin{center}
 \includegraphics[width=.8\textwidth]{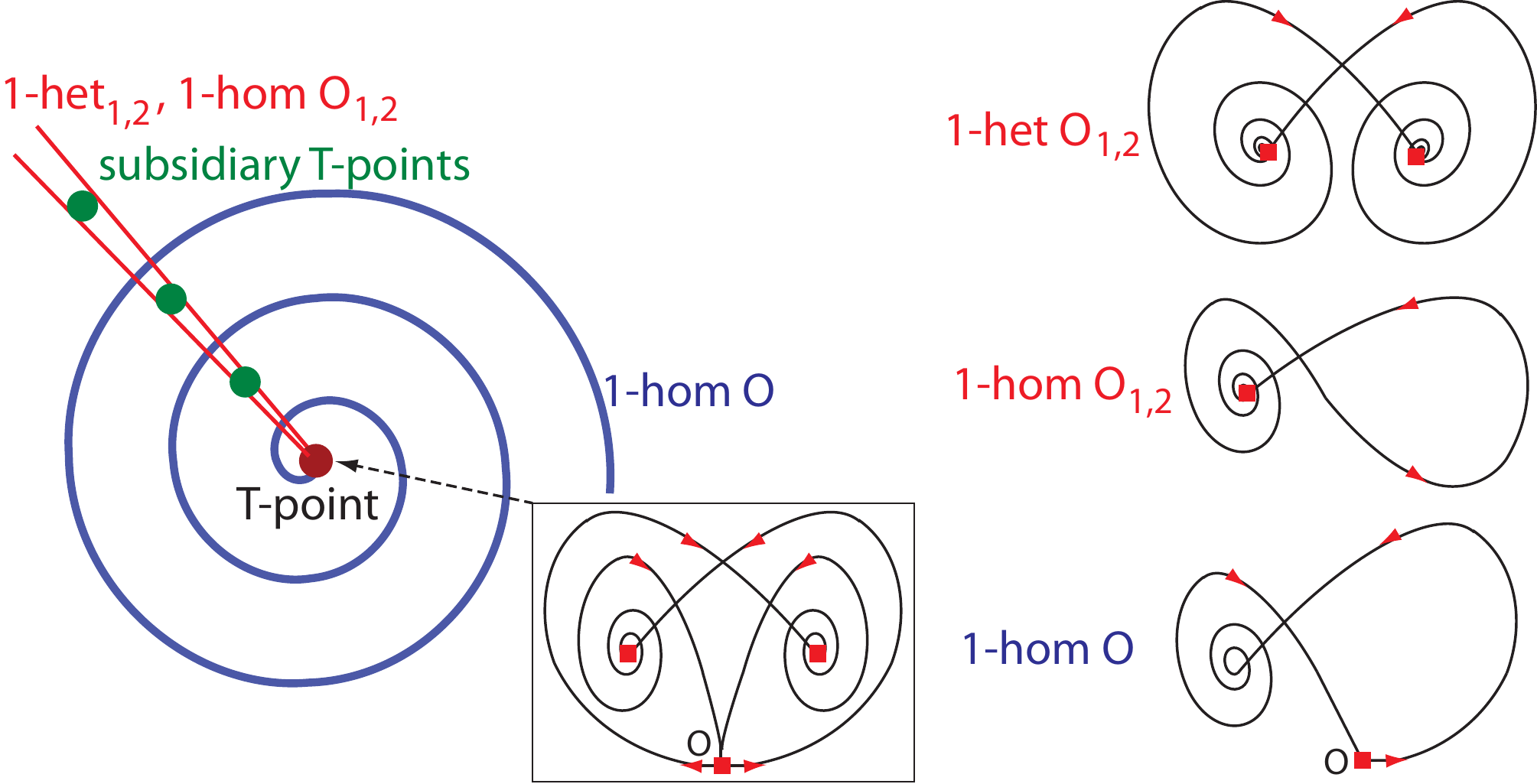}
\caption{Caricature of the bifurcation unfolding of the ordinary T-point for symmetric Lorenz-like systems with a closed heteroclinic
connection evolving both saddle-foci and the saddle.  Point out that with each revolution approaching the T-point along the curve (1-hom $O$), the number of turns of the one-dimensional separatrix of the saddle, $O$, around the saddle-focus increases by one in the homoclinic loop and becomes
infinite at the T-point.}\label{figTpoint}
 \end{center}
\end{figure}

Figure \ref{figTpoint} sketches an idea of the structure of the bifurcation unfolding near an ordinary T-point in the parameter plane in a symmetric system \cite{BYK93,GS86}; various T-points configurations for other heteroclinic connections were examined in detail in \cite{BYK93}. Here, the heteroclinic connection is formed by a pair of symmetric saddle-foci and a saddle whose one-dimensional stable (incoming) and unstable (outgoing), respectively, separatrices merge only at the codimension-two T-point in the bi-parametric space of the Lorenz model. It follows from the theoretical analysis that the unfolding of this bifurcation includes three main curves ending at the T-point: a spiraling bifurcation curve, (1-hom $O$) corresponding to two simultaneous  (due to the symmetry) homoclinic loops of the saddle, a curve, (1-hom $O_{1,2}$),  corresponding to two simultaneous homoclinic loops of the saddle-foci,  and another codimension-one bifurcation curve, (1-het $O_{1,2}$),
corresponding to a heteroclinic correction between both saddle-foci. In addition, the unfolding includes infinitely many subsidiary T-points
in between  1-hom $O_{1,2}$ and 1-het $O_{1,2}$.

Despite a rather overwhelming number of studies reporting the occurrence of various spiral structures, there is yet unproportionately
little known about construction details and generality of underlying bifurcation scenarios giving rise to such patterns.
In this paper we introduce a novel computational toolkit capitalizing on the idea of the symbolic representation for the dynamics of Lorenz-like
systems that employs the kneading invariants \cite{MT88}. We will then show how the toolkit can be used for detecting various structures in
bi-parametric scans of such systems.  It is our intention to enhance further the technique thus allowing for systematic studies of the stunning complexity and universality of spiral structures in models of diverse dynamics and origins.

The paper is organized as follows: in Section \ref{sec:2} we review the homoclinic bifurcation theory and discuss the routes to chaos
and the formation stages of the strange attractor in the Lorenz model; in Section \ref{sec:3} we introduce basics of kneading theory; in Section \ref{sec:4} we demonstrate of the computational technique using kneading invariants to reveal hidden structures of biparametric chaos in
the iconic Lorenz model; in Section \ref{sec:5} we apply the technique to uncover colorfulness of the bifurcation diagrams of the Shimizu-Morioka model. Finally, In Section \ref{sec:6} the proposed technique will be tested on a 6D model of the optically pumped, far infrared red three-level
laser \cite{Moloney1989,FMH91} to confirm the universality of the patters produced by the deterministic chaos in the Lorenz like systems.

\section{Homoclinic bifurcations in systems with the Lorenz attractor}
\label{sec:2}

The strange chaotic attractor in the Lorenz equation from hydrodynamics has become  a de-facto proof of deterministic chaos.
The butterfly-shaped image of the iconic Lorenz attractor, shown in Fig.~\ref{fig1l}, has become
the trademark of Chaos theory and Dynamical Systems. This theory elaborates on complex trajectory behaviors in nonlinear systems
from mathematics, physics, life sciences, finance, etc. Universality of the methods along with bifurcation tools has made them spread
wide and deep across all other disciplines of the modern science.

The Lorenz equation \cite{LO63} is a system of three differential equations:
\begin{equation}
\dot{x} = -\sigma(x-y), \quad \dot{y} = r\,x -y -xz, \quad \dot{z} = -b z+ xy,\label{lorenz}
\end{equation}
with three positive bifurcation parameters: $\sigma$ being the Prandtl number quantifying the viscosity
of the fluid, $b$ being a positive constant of magnitude order 1 which originates from
the nonlinearity of the Boussinesq equation, and $r$ being a Reynolds number that characterizes the fluid dynamics.
Notice that Eqs.~(\ref{lorenz}) are  $\mathbb{Z}_2$-symmetric, i.e. $(x,y,z) \leftrightarrow (-x,-y,z)$ \cite{SPA82}.

\subsection{Uni-parametric cut through the Lorenz equation}

\begin{figure}[hbt!]
\begin{center}
\includegraphics[width=.7\textwidth]{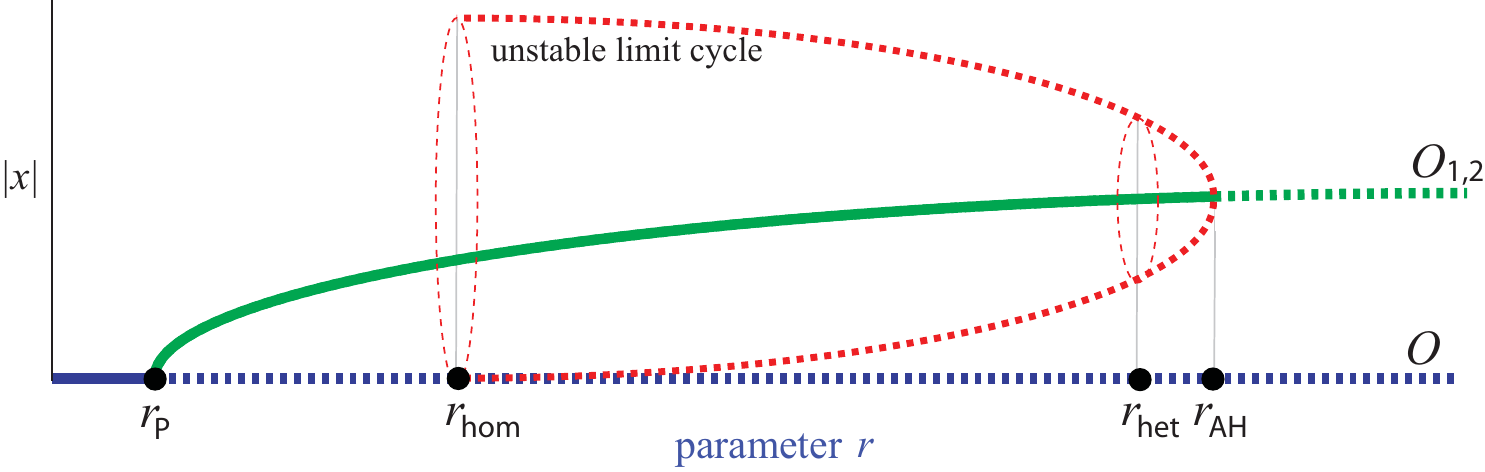}
\caption{Sketch of the uni-parametric bifurcation diagram for the Lorenz equation at $\sigma=10$ and $b=8/3$: plotted
are the coordinates, $|x|$, of the limit trajectories (equilibria, periodic and homoclinic orbits)
against the bifurcation parameter $r$.}\label{figure1}
 \end{center}
\end{figure}

An exploration of primary bifurcations in the Lorenz equation begins with a single-parameter examination of the dynamics, where
$r$ serves as the bifurcation parameter increasing from laminar to weekly turbulent magnitudes around 25,
while the two other parameters are set to the original Saltzman values: $\sigma=10$ (for the water, and $\sigma=1$ for the air)
and $b=8/3$. This would give a uni-parametric cut through the Lorenz equation that was originally explored in these independent studies
\cite{ABS77,KY79} (see Fig.~ \ref{figure1}):
\begin{itemize}
\item For $r<1$  the only equilibrium state $O(0,0,0)$ at the origin is a global attractor in the 3D phase space of the Lorenz equation.
\item This equilibrium state undergoes a pitch-fork bifurcation at $r=r_{\rm P}=1$,  and for $r>1$ becomes a saddle so that the stability is
transferred to two symmetric stable foci.
\item At $r=r_{\rm hom}\approx 13.9162$ the unstable separatrices of the saddle return to the origin thus forming a homoclinic butterfly.
This causes a `` homoclinic explosion'' in the phase space of the model that becomes filled in at once with countably many saddle periodic orbits
that would further compose the skeleton of the Lorenz attractor.
\item For $r_{\rm hom}< r < r_{\rm het}\approx 24.0579$ the model exhibits transient chaos with two ultimate attractors: stable foci $O_{1,2}$.
Such transient chaos is associated with a pre-turbulence regime.
\item The value $r=r_{\rm het}$ corresponds to the onset of the Lorenz attractor coexisting with stable foci $O_1$ and $O_2$. The
attraction basins of $O_{1,2}$ are bounded by the 2D cylindrically-shaped stable manifolds of two saddle periodic orbits that have
earlier bifurcated from the homoclinic loops of the saddle at $r_{\rm hom}\approx 13.9162$.
\item As $r$ increases to $r_{\rm AH}\approx  24.7368$ the saddle periodic orbits shrink the attraction basins and
collapse onto the stable foci  $O_{1,2}$ through a subcritical Andronov-Hopf bifurcation.
\item For $r_{AH} < r <r_{\rm T} \approx  31$ the Lorenz equation possesses a genuinely strange chaotic attractor, known
as the Lorenz attractor, containing no stable orbits.
\end{itemize}

\subsection{Canonical 2D bifurcation diagram of the Lorenz equation}

\begin{figure*}[ht!]
 \begin{center}
 \includegraphics[width=1.\textwidth]{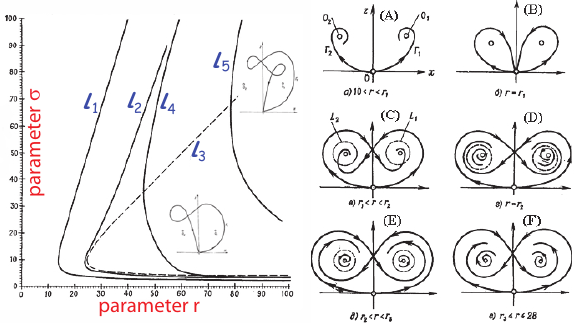}
\caption{Left panel: the $(r,\, \sigma)$-parameter plane depicting the primary bifurcation curves and the stages of the
formation of the Lorenz attractor that are sketched in the right panel. Curve $l_1$ corresponding to the primary homoclinic butterfly shown in (B) ; $l_2$ being the first boundary of the existence region of the Lorenz attractor: stages (C)-(D)-(E); $l_3$ corresponding to a subcritical  Andronov-Hopf bifurcation; and  $l_4$ and $l_5$ (F), corresponding to the homoclinic loops (depicted in in the insets) with symbolic kneadings $\{ +1,-1, 0 \}$  and $\{+1,-1,+1,0\}$, respectively.
Courtesy of \cite{LP1980}.}
\label{fig2l}
 \end{center}
\end{figure*}

The pilot study of the dynamics of the Lorenz equation needs to be further enhanced by the bi-parametric examination of the model, including
at large parameter values \cite{BS07,BS09}.  We will start off with the canonical bifurcation diagram, shown in the
left panel of Fig.~\ref{fig2l} (courtesy of \cite{LP1980}), of the Lorenz equation that depicts the basic bifurcation curves in the
$(r,\sigma)$-parameter plane with fixed $b=8/3$.
The right panel of Fig.~\ref{fig2l} sketches en route fragments of the formation of the Lorenz attractor on the pathway, $\sigma=10$ \cite{ABS77,KY79}
through the bifurcation curves. For $r<1$, Eq.~(\ref{lorenz}) has a single stable equilibrium state at the origin. This equilibrium state
undergoes a pitch-fork bifurcation at $r=1$, so that for $r>1$  the origin becomes a saddle, $O$, of the
topological type (2,1) due to the characteristic exponents $\lambda_3<\lambda_2<0<\lambda_1$.
The saddle has a 1D unstable manifold, $W_O^u$ is made of $O$ itself and a pair of 1D unstable separatrices, $\Gamma_1$ and $\Gamma_2$ (due to $\lambda_1$) entering the saddle as $t \to -\infty$, and a 2D stable manifold, $W_O^s$, containing the leading (due to $\lambda_2$) invariant $z$-axis;
the eigenvector due to $\lambda_3$ determines the non-leading or strongly stable direction, $W^{ss}_{O}$ in $W_O^s$.
After the pitch-fork bifurcation, the  separatrices of the saddle tend to two symmetric attractors -- equilibrium states, $O_{1,2}(x=y=\pm \sqrt{b(r-1)}, z=r-1)$ (Fig.~\ref{fig2l}(A)) that become the global attractor for all trajectories in the phase space of the Lorenz equation other than in $W_O^s$.

A homoclinic butterfly bifurcation occurs in the Lorenz equation when both separatrices, $\Gamma_1$ and $\Gamma_2$,
of the saddle come back to the origin along the $z$-axis (Fig.~\ref{fig2l}(B)). In virtue of the  symmetry of the Lorenz equation, such homoclinic loops are always formed in pairs,  and therefore  constitute bifurcations of codimension-one, in general.  The bifurcation of the homoclinic
butterfly takes place on the curve $l_1$ in the $(r,\sigma)$-parameter plane. Bifurcations of the separatrices of the
saddle at the origin are crucial for the Lorenz attractor.

The very first homoclinic butterfly made of two separatrices looping a single round about the equilibrium states $O_{1,2}$,
causes the homoclinic explosion in the phase space of the Lorenz equation. This  bifurcation
gives rise to a onset of countably many saddle periodic orbits that forming an unstable chaotic (saddle) set , which is
not an attractor yet. For this explosion to happen, the so-called {\em saddle value}  $S=\lambda_1 + \lambda_2$, which
is the  sum of the leading characteristic exponents of the saddle, must be positive; alternatively,  the saddle index $\nu=|\lambda_2|/\lambda_1<1$.
Otherwise, if $S<0$ ($\nu>1$) the homoclinic butterfly  produces a symmetric figure-8 periodic
orbit in the aftermath of the gluing bifurcation through which two stable periodic orbits merge after flowing into the separatrix loops.
L.~Shilnikov~\cite{LP68} pointed out two more conditions, in addition to the primary one (1): $\sigma \ne 0$, or $\nu \ne 1$,  needed for the separatrix loop of the saddle in $\mathbb{R}^{3}$ and higher dimensions to produce a single saddle periodic orbit; they are: (2) $\Gamma \in W^{ss}$, i.e. the separatrix comes back to the saddle along the leading direction, (3) the so-called separatrix value ($A$ in the mapping Eq. (\ref{map}) below) does not vanish, its sign determines whether  the separatrix loop is oriented or twisted, and hence the stable manifolds of the saddle periodic orbit are homeomorphic to a cylinder or a M\"obius band.
Otherwise, he predicted \cite{LP81} that the Lorenz attractor could be born right near such codimension-2 bifurcations, termed resonant saddle, orbit- and inclination-flip, correspondingly \cite{SSTC01}, as it occurs in the Shimizu-Morioka and similar models \cite{ALS86,Rob89,Ry90} below.

Out of many saddle periodic orbits, which exploded the phase space of the Lorenz equation, two are the special,  $L_{1,2}$, ones as they demarcate the
thresholds of the ``interior" of the chaotic unstable set (Fig.~\ref{fig2l}(C)).
After the homoclinic butterfly bifurcation, in the region between the bifurcation curves $l_1$ and $l_2$ in the ($r,\sigma$)-parameter plane,
the separatrices $\Gamma_1$ and $\Gamma_2$ of the saddle switch the targets: now the right/left separatrix tends the left/right stable focus $O_{2,1}$.

In order for the unstable chaotic set to become the Lorenz attractor, it must become invariant, i.e. a closed set containing
all $\alpha$-limit orbits, and hence no loose of trajectories escaping to stable foci $O_{1,2}$. This occurs on the
bifurcation curve, $l_2$, in the parameter space (Fig.~\ref{fig2l}(D)).
To the right of the curve, the basin of the Lorenz attractor is shielded  away from those of the stable equilibrium states by the
2D cylinder-shaped stable manifolds of the two ``threshold" saddle orbits, $L_{1,2}$ that have simultaneously emerged from both
separatrix loops, $\Gamma_{1,2}$ at the homoclinic explosion on the curve $l_1$ in the parameter plane.

\begin{figure*}[ht!]
 \begin{center}
 \includegraphics[width=.6\textwidth]{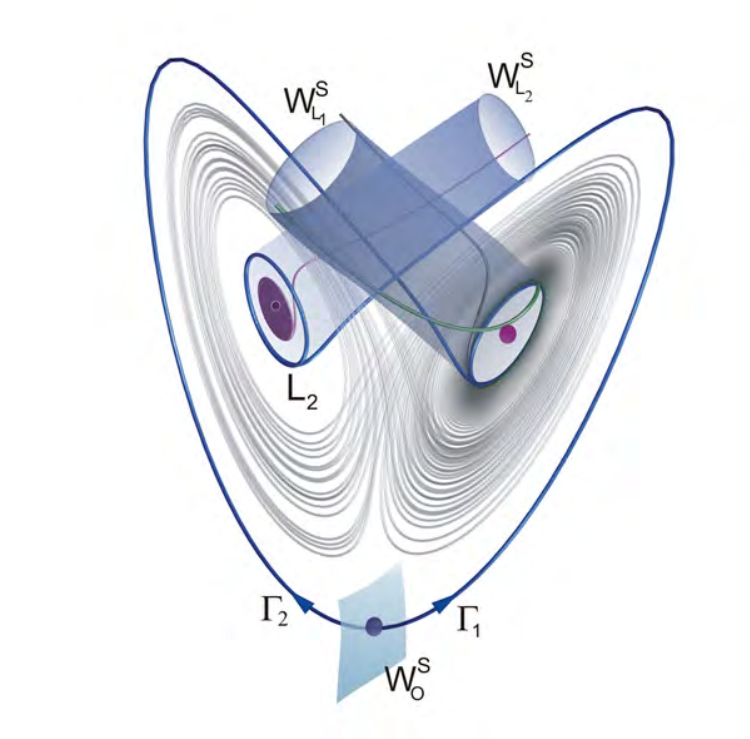}
\caption{3D version of Inset~\ref{fig2l}(D).
The birth of the Lorenz attractor (grey): the attraction basins of the stable foci (purple dots) being blocked away from the extreme
separatrices (blue orbits), $\Gamma_{1,2}$, of the saddle, $O$, at the origin and other trajectories on the Lorenz attractor
by the cylinder-shaped 2D stable manifolds $W^s_{L_{1,2}}$ (dark blue) of the saddle periodic orbits $L_{2,1}$ in the phase space of the
Lorenz equation.}
\label{fig2ln}
 \end{center}
\end{figure*}

As one moves further to the right in the parameter plane, the saddle orbits, $L_{1,2}$, keep narrowing the attraction basins of the foci $O_{1,2}$,
and on the bifurcation curve $l_3$ they collapse into the stable equilibria. The equilibrium states become saddle-foci of the (1,2)-type through a subcritical Andronov-Hopf bifurcation \cite{roshchin78}.
The topological (1,2)-type means that each saddle-focus has 2D unstable and 1D stable manifolds; the latter is formed by two incoming separatrices.
Some local properties of the saddle-foci can be revealed without evaluating their characteristic exponents explicitly. Let
$\lambda_1<0$ stand for the real stable exponent of $O_{1,2}$, and $\lambda_{2,3}$ stand for a complex conjugate pair  such that
$\mathrm{Re} \lambda_{2,3}>0$. Observe that the divergence of the vector field defined by Eqs.~(\ref{lorenz}) is given by $[-\sigma-1-8/3]$,
which equals $\sum_{i=1}^3 \lambda_i=\lambda_1+2\mathrm{Re} \lambda_{2,3}<0$. This implies $\lambda_1+\mathrm{Re} \lambda_{2,3}<0$,
i.e., the complex conjugate pair is closer to the imaginary axis than the real negative exponent, and hence the saddle-foci meet the Shilnikov condition \cite{sh65,LP67,LPALS07}. Therefore, as soon as the saddle-focus possesses a homoclinic loop, such a bifurcation causes the abundance of periodic orbits nearby. Those periodic orbits constantly undergo saddle-node and period doubling bifurcations as the parameters are varied.
Moreover, since the divergence of the vector field of the Lorenz equation is always negative, saddle-node bifurcations give rise to stable periodic orbits near the homoclinic saddle-focus bifurcation.  Under fulfillment  of some global conditions, a single Shilnikov saddle-focus bifurcation can lead to the formation of a spiral or screw-like attractor. However,  a strange attractor due to the Shilnikov saddle-focus in a 3D system with a negative divergence is no genuinely chaotic set in the sense that it contains stable periodic orbits within. Because of that, such a chaotic attractor is called quasi-attractor thus referring to that because besides stable periodic orbits with weak
basins it may have structurally unstable or non-transverse homoclinic orbits   \cite{AS83,LP84,LP97}. Note that systems in higher dimensions can possess
genuinely strange attractors with the Shilnikov loop without stable periodic orbits, the so-called wild chaotic attractors
\cite{LP84,LP97,TLP98,LP02}.

\begin{figure}[hbt!]
 \begin{center}
 \includegraphics[width=.6\textwidth]{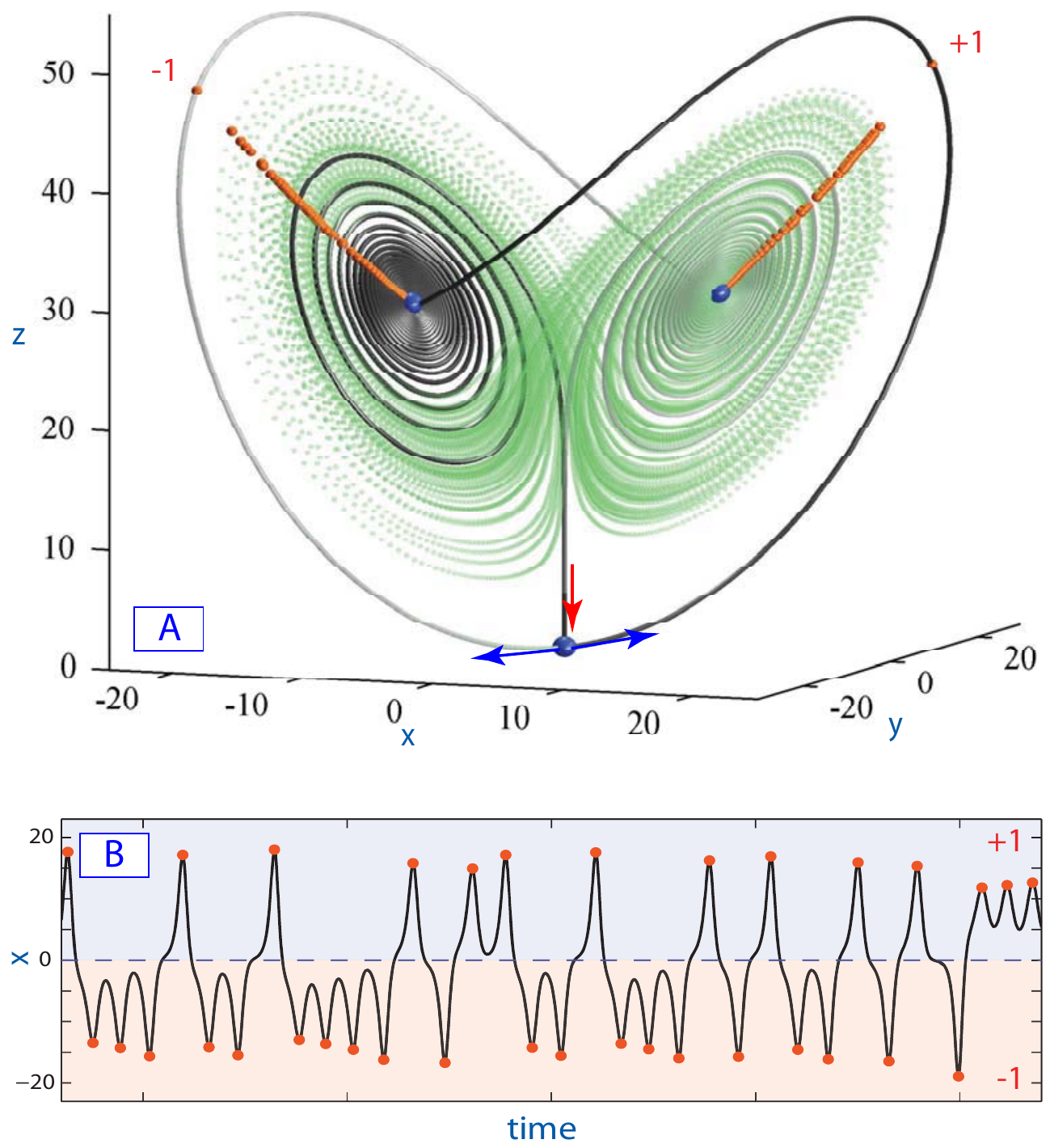}
\caption{(A) Heteroclinic connection (in dark color) between the saddle at the origin and two saddle-foci (blue spheres) being
overlaid with the Lorenz attractor (green light color) on the background at the primary T-point $(r=30.38,\,\sigma=10.2)$.
Orange spheres on the butterfly wings indicating the turning points around the right and left saddle-foci define the kneading sequence entries, $\{\pm 1\}$, respectively. (B) A typical time evolution of the $x$-coordinate of the right separatrix of the saddle.}\label{fig1l}
 \end{center}
\end{figure}

The Lorenz attractor is non-hyperbolic because it includes the singularity at the origin  --  the saddle equilibrium state with an 1D unstable manifold while all other saddle periodic orbits on the attractor have 2D stable and unstable manifolds. Moreover, the manifolds of those orbits in the Lorenz attractor (self) cross transversally in the 3D phase space thereby producing only structurally stable (transverse) homoclinic and heteroclinic trajectories. This condition imperative for the Lorenz attractor will not always hold for larger parameter values and lead to
homoclinic tangencies of the manifolds which are followed by saddle-node bifurcations in the Newhouse regions \cite{GTS93,GTS96} of the parameter plane
of the system. Thus, in order for the Lorenz attractor to be strange and chaotic with no stable orbits, it must not include the [homoclinic] saddle-foci, $O_{1,2}$, as well as contain only structurally stable homoclinic orbits due to transverse intersections of the manifolds of saddle periodic orbits.


\section{Kneading invariants}
\label{sec:3}

Chaos can be quantified by several means. One customary way is through the evaluation of the topological entropy.
The greater the value of topological entropy, the more developed and unpredictable the chaotic dynamics become. Another practical approach
for measuring chaos in simulations capitalizes on evaluations of the largest (positive) Lyapunov exponent of a long yet
finite-time transient on the chaotic attractor.

After the stable foci have lost the stability through the subcritical Andronov-Hopf bifurcation, the Lorenz equation exhibits the  strange attractor of the iconic butterfly shape. The wings of the butterfly are marked with two symmetric eyes containing the saddle-foci isolated from  the trajectories of the Lorenz attractor. This attractor is structurally unstable \cite{GW79,ABS83} as it undergoes bifurcations constantly as the parameters of the Lorenz equation are varied. The primary cause of structural and dynamical instability of chaos in the Lorenz equation and similar models is the singularity at the origin -- the saddle with two one-dimensional outgoing separatrices. Both separatrices fill in densely two spatially symmetric, [$(x,y,z) \leftrightarrow (-x,-y,z)$], wings of the Lorenz attractor in the 3D phase space (see Fig.~\ref{fig1l}).
The Lorenz attractor  undergoes a homoclinic bifurcation when the separatrices of the saddle change the flip-flop pattern of switching between the  butterfly wings centered around the saddle-foci. At such a change, the separatrices comes back to the saddle thereby causing additional homoclinic explosions in phase space \cite{ABS77,KY79}.

The time progression of the ``right" (or symmetrical ``left") separatrix  of the origin can be described geometrically and categorized in terms of the number of flip-flops around the equilibrium states $O_{1}$ and  $O_{2}$ in the 3D phase space of the Lorenz equation (Fig.~\ref{fig1l}). Or, alternatively, can be reduced to  the time-evolution of the $x$-coordinate of the separatrix,  as shown in panel~B of Fig.~\ref{fig1l}. The sign-alternation of the $x$-coordinate suggests the introduction of a $\{\pm 1\}$-based alphabet to be employed for the symbolic description of the separatrix. Namely, whenever the right separatrix turns around $O_1$ or $O_2$, we write down $+1$ or $-1$, respectively. For example, the time series shown in panel~B generates the following kneading sequence starting with
$\{ +1, -1,-1,-1,+1,-1,-1,+1,-1, \ldots \} $ etc.

\begin{figure*}[!htb]
\begin{center}
\includegraphics[width=.4\textwidth]{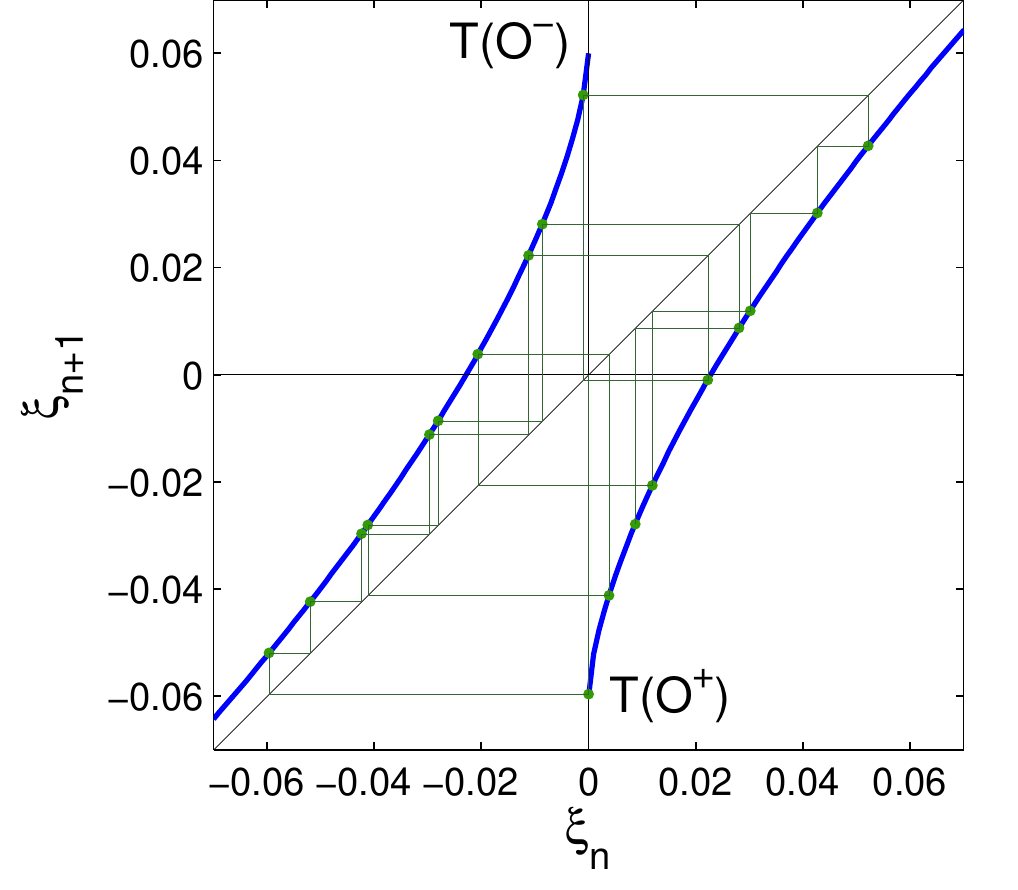}~\includegraphics[width=.4\textwidth]{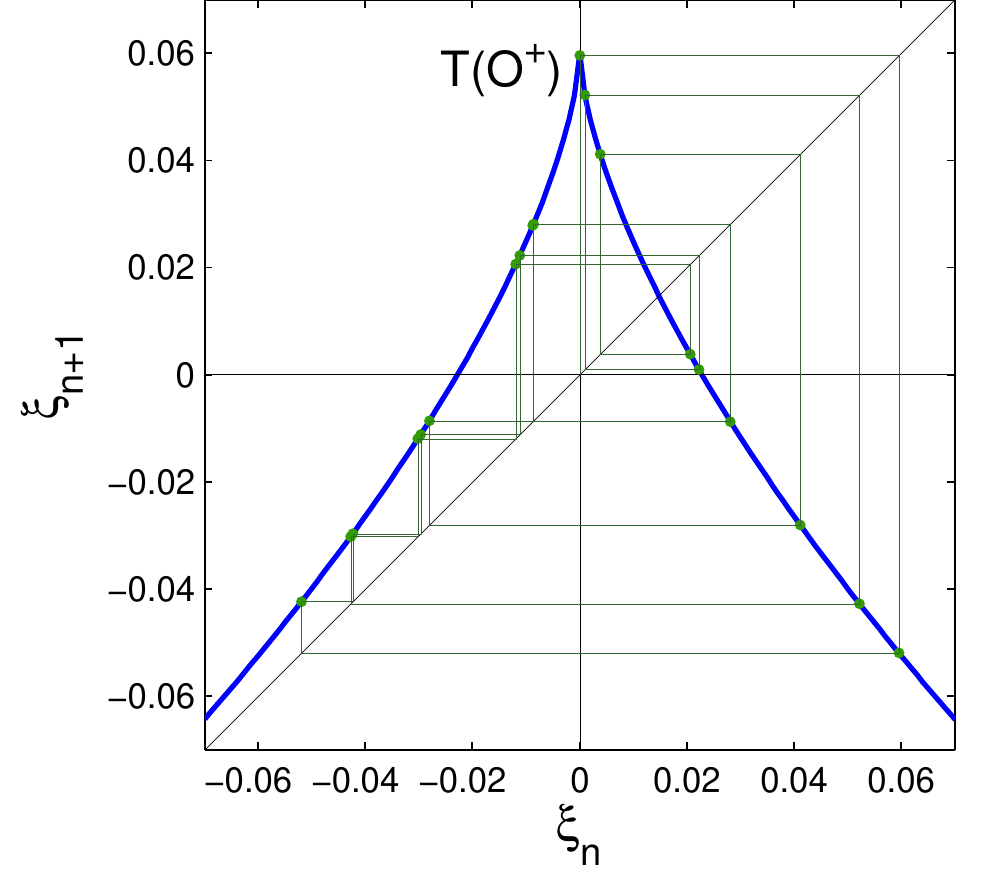}
\caption{\label{fig5} (A) 1D Lorenz mapping (geometric model) with the discontinuity corresponding to the saddle at the origin in the 3D phase space
of the Lorenz equation. Shown are the forward iterates of the ``right'' separatrix, $O^+$, of the discontinuity point. Iterates on the right, $x >0$, and left, $x<0$, branches of the mapping are assigned to kneading symbols of $+1$, and $-1$, respectively. Here $\nu=0.65$, $A=0.7$ and $\mu=-0.06$ in Eq.~(\ref{map}). (B) Alterative cusp-shaped mapping as it would be for the $z$-variable of the separatrix at the turning  points, given by $z_{\rm max}^\prime(t)=0$, on the Lorenz attractor.}\label{1dmapping}
 \end{center}
\end{figure*}

In what follows we will introduce and demonstrate a new computational toolkit for the analysis of chaos in the Lorenz-like models. The toolkit
is inspired by the idea of kneading invariants introduced in \cite{MT88}. A kneading invariant is a quantity that is intended to uniquely describe
 the complex dynamics of the system that admit a symbolic description using two symbols, here $+1$ and $-1$. The kneading invariant is supposed to depend monotonically on the governing parameter so that any two systems can be compared and differentiated, or equivalently,
ordered in terms of  $>$ and  $<$, by the kneadings.  Two systems with the Lorenz attractors   are topologically conjugate when they share the
same kneading invariant, see \cite{GW79,R78,Malkin91,GS93,TW93} and the references therein. Moreover, for the Lorenz-like systems, the topological entropy can be evaluated though the kneading invariants by reducing consideration to piecewise monotone mappings of a closed interval  \cite{GH96,Malkin2003}.

Such mappings  are closely related to the geometric models of the Lorenz attractor which are 1D and 2D Poincar\'e return mappings
defined on some cross-section transverse to trajectories of the Lorenz attractor. The basic idea behind either geometric model capitalizes on
extended properties of the local Poincar\'e mapping near a homoclinic butterfly of the saddle \cite{GW79,ABS83,SSTC01}. The mapping is assumed to
meet a few other conditions related to the global properties of the flow far from the homoclinic butterfly. Such a 1D constrained mapping shown in Fig.~\ref{1dmapping}(A) can be written as:
\begin{equation}
\label{map}
T:~~\xi_{n+1} = \left ( \mu + A |\xi_{n}|^{\nu} \right ) \cdot \mbox{sign}(\xi_n),
\end{equation}
here $\nu={|\lambda_2|}/{\lambda_1}<1$ is the saddle index, $\mu$ controls the distance between the returning separatrices, $\Gamma_{2,1}$,
and the saddle, and $A$ is a scalar whose sign determines whether the homoclinic loops at $\mu=0$ are oriented if $A>0$ , or twisted when $A<0$,
see \cite{SSTC01} for more details.

Loosely speaking, this geometric model (\ref{map}) should have no critical point on both branches, and moreover possess a property of
strong stretching with a rate of expansion more than $\sqrt{2}$ \cite{ABS83}. This would guarantee that the Lorenz
attractor will be densely filled in by the forward iterates of the separatrices, $O^{\pm}$ with no holes -- lacunas containing
isolated periodic orbits inside, stable or not. Strong stretching would also imply a monotone dependence of the kneading invariants on a
governing parameter.

In a symmetric system with the Lorenz attractor, the kneading invariant is assigned to quantify the symbolic description of either separatrix;
in the asymmetric case one should consider two kneading invariants for both separatrices of the saddle.
Thus, in respect it reflects quantitatively a qualitative change in the separatrix behavior, such as flip-flopping patterns,
as the parameter of the system is changed.

By construction, kneading invariants are proposed to serve as moduli of the topological equivalence that are employed to compare or contrast between any two Lorenz attractors or, equivalently, any two Lorenz-like systems. Due to the symmetry of the Lorenz mapping $\xi_{n+1}= T(\xi_n)=T^n(\xi_0)$ from Eq.~(\ref{map}), forward iterates of the right separatrix, $O^+$, of the discontinuity point (resp. the saddle) are detected to generate
a {\em kneading sequence} $\{ \kappa_n(O^+) \} $ defined
by the following rule:
\begin{equation}
\kappa_n (O^+) =  \left\{
\begin{array}{cc}
  +1, &  \mbox{if~~} T^n(O^+) > 0,  \\
-1, &  \mbox{if~~} T^n(O^+) < 0, \\
~0, & \mbox{if~~} T^n(O^+) = 0;
\end{array}\right.
\end{equation}
here $T^n(O^+)$ is the $n$-th iterate of the right separatrix $O^+$ of the origin. The condition  $T^n(O^+) = 0$ is interpreted as
a homoclinic loop, i.e. the separatrix returns to the origin after $n$ steps.

\begin{figure*}[!htb]
\begin{center}
\includegraphics[width=0.8\textwidth]{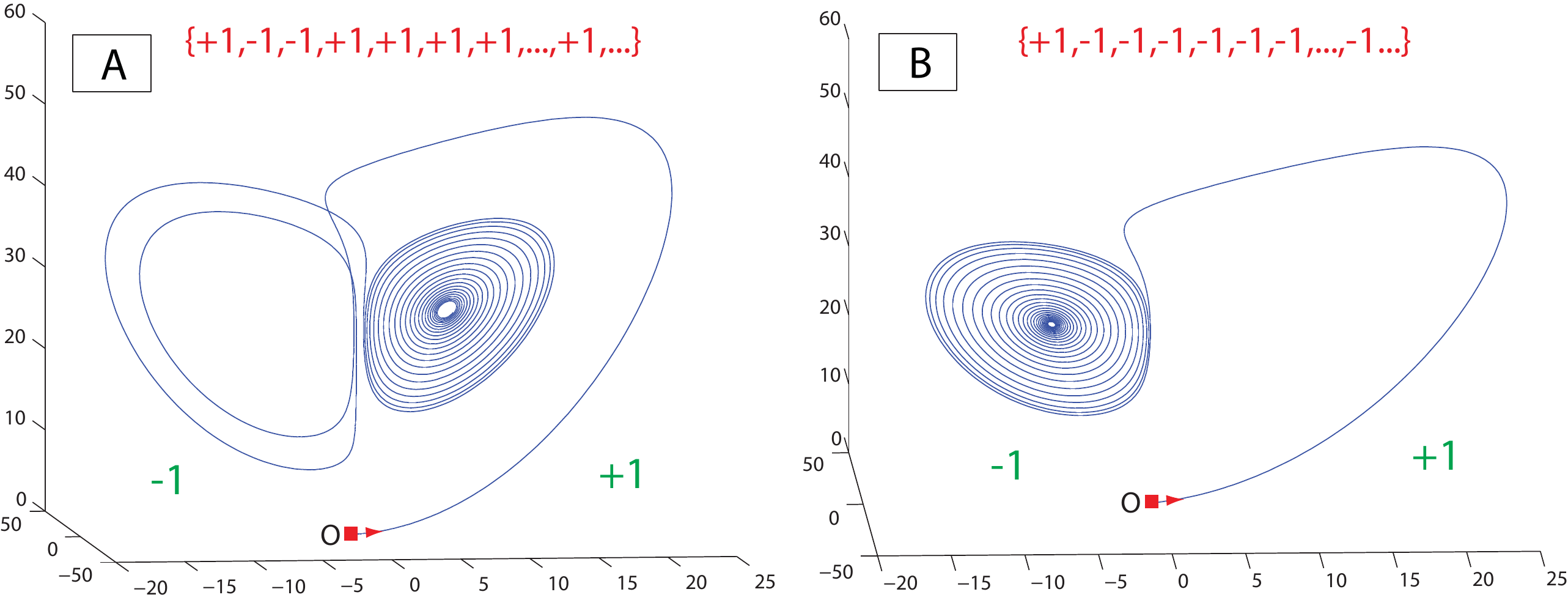}
\caption{\label{figtwo}  Truncated kneading sequences generated by the right outgoing separatrix of the saddle at the origin in the Lorenz
equation at two distinct values of the parameters.}
 \end{center}
\end{figure*}

The  kneading invariant for the separatrix is defined in the form of a formal power series:
\begin{equation}
P(q) = \sum_{n=0}^\infty {\kappa}_{n}\,q^n \label{fs}.
\end{equation}
Setting $q\in (0,\,1)$ make the series~(\ref{fs}) convergent. The smallest zero, $q^*$, if any, of the graph of (\ref{fs}) in the interval $q\in (0,\,1)$ yields the topological entropy, $h(T) = \ln(1/q^*)$, of the 1D mapping~(\ref{map}).

Let us next draw a parallelism between the geometric model and the Lorenz equation: the kneading sequence $\{{\kappa}_{n}\}$ comprised of only $+1$s of the mapping~(\ref{map}) corresponds to the right separatrix converging to the stable equilibrium state, $O_1$ (or possibly, a periodic orbit with $x(t)>0$).
The corresponding kneading invariant is maximized at $\{P_{\rm max}(q)\}=1/(1-q)$. When the right separatrix converges to an
$\omega$-limit set with $x(t)<0$, like the left stable focus, $O_{2}$ then the kneading invariant is given by $\{P_{\rm min}(q)\}=1-q/(1-q)$ because the first entry $+1$ in the kneading sequence is followed by infinite $-1$s. Thus, $\left [\{P_{\rm min}(q)\},\,\{P_{\rm max}(q)\} \right ]$ yield
the range of the kneading invariant values; for instance $\left [\{P_{\rm min}(1/2)\}=0,\,\{P_{\rm max}(1/2)\}=2 \right ]$.
Two samples of the separatrix pathways shown in Fig. \ref{figtwo} generating the following kneading invariants
$$
\begin{array}{l}  P_{A}(1/2)=+1-1/2-1/4+1/8+1/16+1/32+1/64 \ldots +1/2^n \ldots = 1/2,\\
P_B(1/2)= +1-1/2-1/4-1/8-1/16-1/32-1/64  \ldots -1/2^n \ldots = 0,
\end{array}
$$
illustrate the parallelism.

\begin{figure*}[p]
 \begin{center}
 \includegraphics[width=0.92\textwidth]{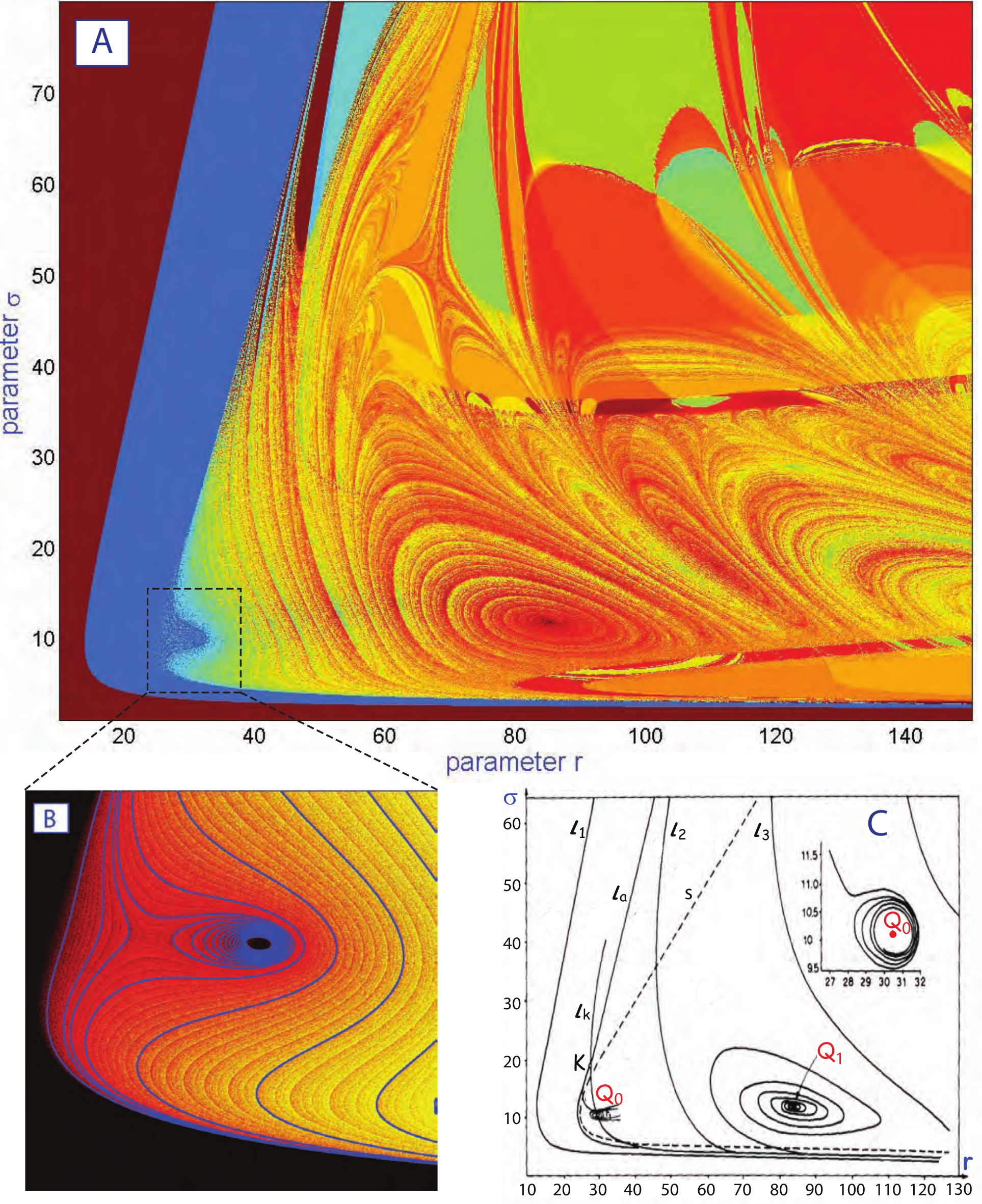}
\caption{(A) Kneading-based bi-parametric scan revealing multiple T-points and saddles that organize globally
complex chaotic dynamics of the Lorenz equation in the $(r,\sigma)$ parameter plane.
Solid-color regions associated with constant values of the kneading invariant correspond to simple
dynamics dominated by stable equilibria or stable periodic orbits. The borderline between
the brown and blue region corresponds to the bifurcation curve of the homoclinic butterfly. The border line between
the blue and yellow-reddish region corresponds to the formation of the Lorenz attractor (below $\sigma \simeq 50$).
 (B) Zoom of the vicinity of the primary T-point at ($r=30.4,\sigma=10.2$) to which
a homoclinic bifurcation curve spirals onto. Data for the homoclinic curves (in blue) are courtesy of Yu. Kuznetsov. (C)
Original bifurcation diagram of the Lorenz equation depicting the two detected T-points and primary
homoclinic bifurcation curves; courtesy of \cite{BYK93}}.
\label{fig3l}
\end{center}
\end{figure*}

To conclude this section we point out that there is another approach for constructing 1D return mappings through the evolution of the
$z$-variable of the separatrix of the Lorenz equation. The mapping generated by the turning points where $z_{\rm max}^\prime(t)=0$ on the
attractor has a distinct cusp-shaped form depicted in Fig.~\ref{1dmapping}(B). The point corresponding to the cusp is used for the initiation of the kneading sequence. Due to its unimodal graph with two, increasing and decreasing, segments  the corresponding formal power series is then defined as follows:
\begin{equation}
\widetilde{P}(q) = \sum_{n=0}^\infty \widetilde{\kappa}_{n}\,q^n, \quad \mbox{where}\quad \widetilde{\kappa}_{n}= \prod_{i=0}^n{\kappa}_{i},  \label{fs2}
\end{equation}
i.e. $\widetilde{\kappa}_{n}={\kappa}_{n} \cdot \widetilde{\kappa}_{n-1}$.


\section{Kneading scanning of the Lorenz equation}
\label{sec:4}

In this section we carry the concept of the kneading invariants over to numerical studies of fine structures of chaos
in the Lorenz equation. For sake of simplicity in this pilot phase, we employ a rough idea for defining the kneading sequences of $+1$s and $-1$s
that relies on whether the right separatrix of the saddle makes a revolution around the right equilibrium state, $O_1$, or the left one, $O_2$,
respectively, in the ($x,z$)-projection of the 3D phase space. One can utilize even a simpler approach where the sign of the current kneading entry in the sequence is determined by the sign of the $x$-coordinate of the separatrix  at the off-lying turning points, $\max|x(t^*)|$ (on the butterfly wings), given by $x'(t^*)=0$, see the trace in Fig.~\ref{fig1l}(B). There are other alternative ways for defining kneading entries, involving cross-sections, $z'(t)$ etc, that are not free of certain limitations either. In the future we plan to enhance the current kneading algorithms by utilizing the incoming 1D separatrices of the saddle-foci and finding the winding numbers around them instead.

In this computational study of the Lorenz equation and two other models below, we consider partial kneading power series truncated after the first
50 entries: $P_{50}(q) = \sum_{n=0}^{50} {\kappa}_{n}\,q^n $.
The choice of the number of entries is not motivated by numerical precision, but by simplicity, as well as by a resolution of the bitmap
mappings for the bi-parametric scans of the models. One has also to figure the proper value of $q$: setting it too small makes the convergence
 fast so that the tail of the series has a little significance and hence does not differentiate the fine dynamics of the Lorenz equation on longer time scales.

At the first stage of the routine, we perform a bi-parametric scan of the Lorenz equation within a specific range in the $(r,\,\sigma)$-plane.
The resolution of scans is set by using mesh grids of $[1000 \times 1000]$ equally-distanced points. Next by accurately integrating
the separatrix using Taylor series software \textsc{TIDES}\footnote{freeware \textsc{TIDES} is a versatile numerical ODE solver for integration
of ODEs with an arbitrary precision, especially for chaotic systems.}
 \cite{abbr09,abbr09b,BRAB11}  we identify and record the sequences $\{\kappa_n\}_{50}$ for each point of the grid in the parameter plane. Then we define the bi-parametric mapping: $(r,\sigma) \to  P_{50} (q)$ with some appropriately chosen  $q$,  the value that determines the depth of the scan.
The mapping is then colorized in Matlab by using various built-in non-linear spectra ranging between to $P_{50}^{\rm min}$ and  $P_{50}^{\rm max}$, respectively.  In the mapping, a particular color in the spectrum  is associated with a persistent value of the kneading invariant on a level curve.
Such level curves densely foliate the bi-parametric scans.

\begin{figure}[ht!]
 \begin{center}
 \includegraphics[width=0.4\textwidth]{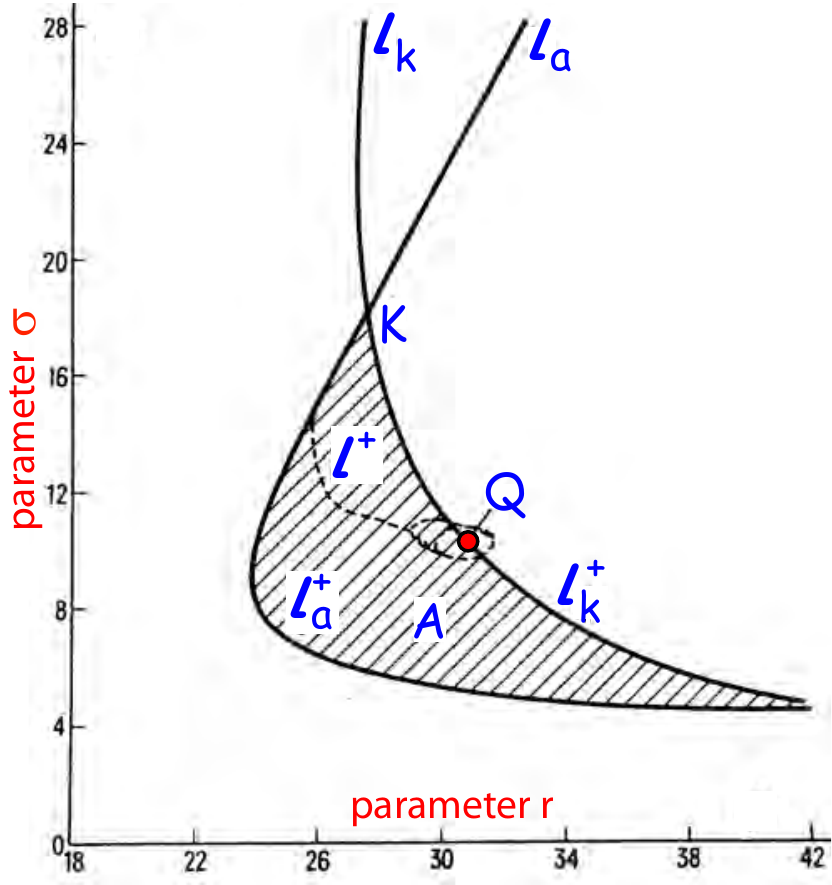}
\caption{The $(r,\,\sigma)$-bifurcation diagram of the Lorenz equation depicting the existence region (shaded) of the Lorenz attractor.
The second bifurcation curve, $l_{k}$, passing through the primary T-point, $Q(r=30.4,\sigma=10.2)$, crosses the first boundary, $l_a$ ($l_2$ in Fig.~\ref{fig2l}), at a point, labeled by $K$, thus closing the existence region. Crossing the branch $l_a$ rightward above the point $K$
does lead to the emergence of the Lorenz attractor: after a long chaotic transient, the separatrix of the saddle will be attracted to either
stable foci, $O_1$ or $O_2$.
Courtesy of \cite{BS92}. }
\label{fig4l}
\end{center}
\end{figure}

Figure~\ref{fig3l} represents the kneading-based color scan of the dynamics  of the Lorenz equation mapped onto a fragment
of the ($r,\,\sigma$)-parameter plane. In the scan, a window of a solid color corresponds to a constant kneading invariant.
In such windows the dynamics of the Lorenz equation are dominated by simple attractors such as stable equilibria or stable periodic orbits
to which the separatrix converge. A quick examination of the kneading definition (\ref{fs}) reveals that the kneading invariant does not vary
at a supercritical Andronov-Hopf bifurcation and a pitch-fork bifurcation describing continuous transitioning between stable symmetric and asymmetric periodic orbits. This holds true for a period-doubling bifurcation too, because the kneading sequence, say $\{(+1,-1,-1)^\infty\}$, inherits the same
block repeated twice $\{(+1,-1,-1,+1,-1,-1)^\infty\}$ in the code for the periodic orbit of a double period and so forth.
While the kneading technique does not detect such safe bifurcation boundaries \cite{SSTC01} having crossed through which the phase point does not
run far from an old attractor to a new one, though it detects dangerous boundaries well, including a generic saddle-node bifurcation,
homoclinic bifurcations, and some other.

A borderline between two solid-color regions corresponds to a bifurcation through a dangerous boundary which is associated with a
jump between the values of the kneading invariant. For example, the borderline in Figure~\ref{fig3l} between the brown region with the kneading sequence
$\{ (+1)^\infty \}$ and the blue region, with the kneading sequence $\{ +1,(-1)^\infty \}$, corresponds
to the primary homoclinic butterfly of the saddle. The second borderline of the blue region corresponds to the onset
of the Lorenz attractor existing on the right from it. One can see that above $\sigma \simeq 50$ this border is adjoined
by windows of solid colors thus indicating that the separatrices start converging to stable equilibria after chaotic transients, long or short.
Indeed, crossing the curve, $l_2$ (or $l_a$ in Fig.~\ref{fig2l}), above $\sigma=18$ does not imply that the Lorenz equation will have a strange attractor, but a chaotic saddle set \cite{BYALS89,BS92}.

What the proposed kneading technique does extraordinary well, compared to the bi-parametric screening based on the finite-time
Lyapunov exponent approach (see below), and what we developed it for, is the detection of bifurcations within the
Lorenz strange attractor. The corresponding yellow-reddish regions in the parameter plane in Fig.~\ref{fig3l} clearly demonstrate the evidence of
the parametric chaos that, like in turbulence, enriched by vortices of multiple T-points.  Panel~B of this figure depicts the kneading
mapping near the left-bottom corner of the bifurcation diagram in panel~A.
In it, the black (blue in panel A) region corresponds to the chaotic saddle dynamics transitioning into the Lorenz attractor
after the mapping color shifts to the yellow-reddish spectrum. Blue curves in panel~B are the bifurcation curves
of some homoclinic orbits with short admissible kneadings. One can see from this panel that the mapping spectrum  is
clearly foliated by the kneading invariant level curves of the colors gradually progressing  from red to yellow. This indicates that the new born
Lorenz attractor, while being structurally unstable and sensitive to parameter variations, persists initially the pseudo-hyperbolic property
because the foliation remains uniform, and transverse to the classical pathway $\sigma=10$ (Fig.~\ref{fig2l}). The homogeneous
foliation starts breaking around a saddle point after which one singled bifurcation curve spirals onto the primary T-point.
Far from this point, the curve corresponds to the formation of the homoclinic loop with the kneading $( 1, -1,-1,-1, 0)$; i.e. the right separatrix
makes one excursion around the saddle-focus $O_1$, followed by three revolutions around the saddle-focus $O_2$, and then returns to the saddle at the origin. While moving along the spiraling curve toward the T-points, the separatrix makes progressively more turns around $O_2$, or more precisely around the 1D incoming separatrix of this saddle-focus. With each incremental turn around  $O_2$, the separatrix comes closer to  $O_2$ while the bifurcation curve becomes one scroll closer to the T-point simultaneously. Due to this feature the T-point $Q_0(r=30.4, \sigma=10.2)$ is called
a Terminal point. The T-point corresponds to the following symbolic sequence $\{ +1, (-1)^\infty\}$. In virtue of the symmetry
of the Lorenz equation, the T-point actually corresponds to a closed heteroclinic connection involving all three saddle-equilibria, see Figs.~\ref{figTpoint} and \ref{fig1l}. The merger of the right (left) separatrix of the saddle $O$  with the incoming separatrix of the saddle-focus
$O_{2}\,(O_1)$, increases the codimension (degeneracy) of this heteroclinic bifurcation to two; note that intersections of the 2D unstable manifolds of the saddle-foci, with the 2D stable manifold of the saddle at the origin are transverse in the 3D phase space in general.
Breaking the 1D heteroclinic connection gives rise to a primary homoclinic orbit to the saddle-focus, as well as to a heteroclinic connection
between both saddle-foci (see the sketch of the bifurcation unfolding of the T-point in Fig.~\ref{figTpoint}).
The corresponding bifurcation curves of these homoclinic and heteroclinic bifurcations originating rightward
from the T-point bound a sector  containing subsidiary T-points \cite{BYK93,GS84}.  Each new T-point produces other
self-similar structures scaled like fractals. Panel~C shows two such identified T-points: primary $Q_0$ and secondary $Q_1\,(r=85.\,\sigma=11.9)$.
The primary codimension-two T-point in the Lorenz equation was originally discovered by Yudovich \cite{PY80}.

\begin{figure}[ht!]
 \begin{center}
\includegraphics[width=.6\textwidth]{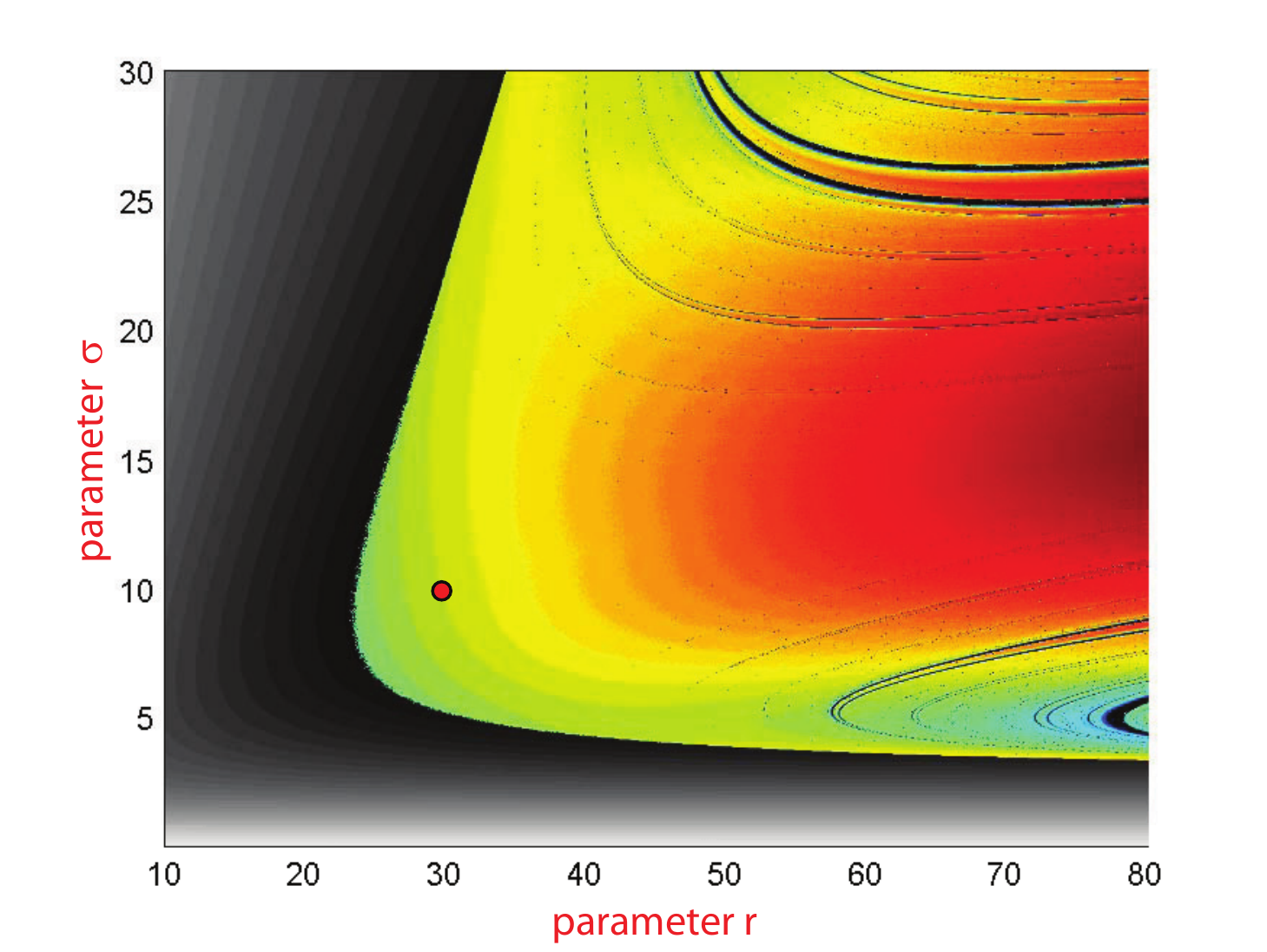}
\caption{\label{fig4la} Finite-time largest-Lyapunov exponent, $L_{\rm max}$, scan of the Lorenz
equation showing no sign of spiral structures in the $(r,\sigma)$-parameter plane.
The dark region corresponds to trivial attractors, where $L_{\rm max} \le 0$,
while the red color indicates $L_{\rm max}>0$ in chaotic regions. The red dot points out the location of the primary T-point.}
 \end{center}
 \end{figure}

As soon as the saddle-foci and their bifurcations become involved in the  dynamics of the Lorenz equation near the primary T-point,
the Lorenz attractor loses the purity of the genuine chaotic attractor that used to have neither stable periodic orbits nor non-transverse homoclinic trajectories. It transforms into a quasi-chaotic attractor with weakly stable orbits of small basins and nontransverse homoclinic orbits. The idea of non-transversality or tangency was employed in \cite{BYALS89,BS92} to numerically identify the second boundary, $l_K$ , in addition to the first one,
$l_a$ ($l_2$ in Fig.~\ref{fig2l}), that bounds the existence region of the Lorenz attractor in the parameter plane, see Fig.~\ref{fig4l}. Note that $l_K$ crosses the initial boundary, $l_a$.  This means that above the intersection point, crossing $l_a$ ($l_2$ in Fig.~\ref{fig2l}) rightwards does not guarantee that the basin of the Lorenz attractor will necessarily be isolated from the basins of the stable foci, $O_{1,2}$ by the stable manifolds
of the saddle periodic orbits, $L_{1,2}$  (Fig.~\ref{fig2ln}). This implies that the separatrices of the saddle will demonstrate chaotic transient behavior, long or short, prior to them converging to $O_{1,2}$, see \cite{ALS86,ALS89,BYALS89,ALS91,BS92,ASHIL93,SST93} for full details.

The other feature of the boundary, $l_K$, is that it passes through the primary T-point, thereby separating the existence region of the Lorenz attractor from bifurcations of the saddle-foci, and consequently from all subsidiary T-points existing on the right from it in the parameter plane.
Here the chaotic dynamics of the Lorenz equation become even wilder and less predictable \cite{LP84,LP97,TLP98,LP02}. The indication to that is panel~A of Fig.~\ref{fig3l}  that reveals, through the kneading scan, a parametric turbulence in the $(r,\sigma)$-parameter plane with fractal explosions in the forms of  multiple spiral structures -- ``tornado eyes"   centered around T-points.
Note that basins of spiraling T-points are separated by corresponding saddles. One can spot self-similar smaller-scale spiral structures
within large-scale ones and so forth. The richness of such fractal structures in the parameter plane results from the synergy of the Lorenz-like dynamics
amplified by chaos induced by the Shilnikov saddle-foci.

To conclude this section we contrast the scans of the Lorenz equation obtained using the proposed kneading technique with the sweeps
based on the evaluation of the largest Lyapunov exponent, $L_{max}$, for the separatrices of the saddle evaluated over a finite time interval \cite{BS07,BS09,BBS11}.  Figure~\ref{fig4la} shows a fragment of the typical bi-parametric sweep of the Lorenz equation: the dark
region at small values of the Reynolds number, $r$, is where $L_{max}$ is negative on the stable foci. The sweep yields
a clear borderline between the regions of the simple and chaotic attractors.  The chaotic region (yellow-reddish) is characterized by a small
positive Lyapunov exponent, variations of which are not significant enough to reveal fine structures, as spiraling T-points.
The method can detect well stability islands in the biparametric diagram which correspond to emergent stable periodic orbits.

\begin{figure*}[h!]
 \begin{center}
\includegraphics[width=1.\textwidth]{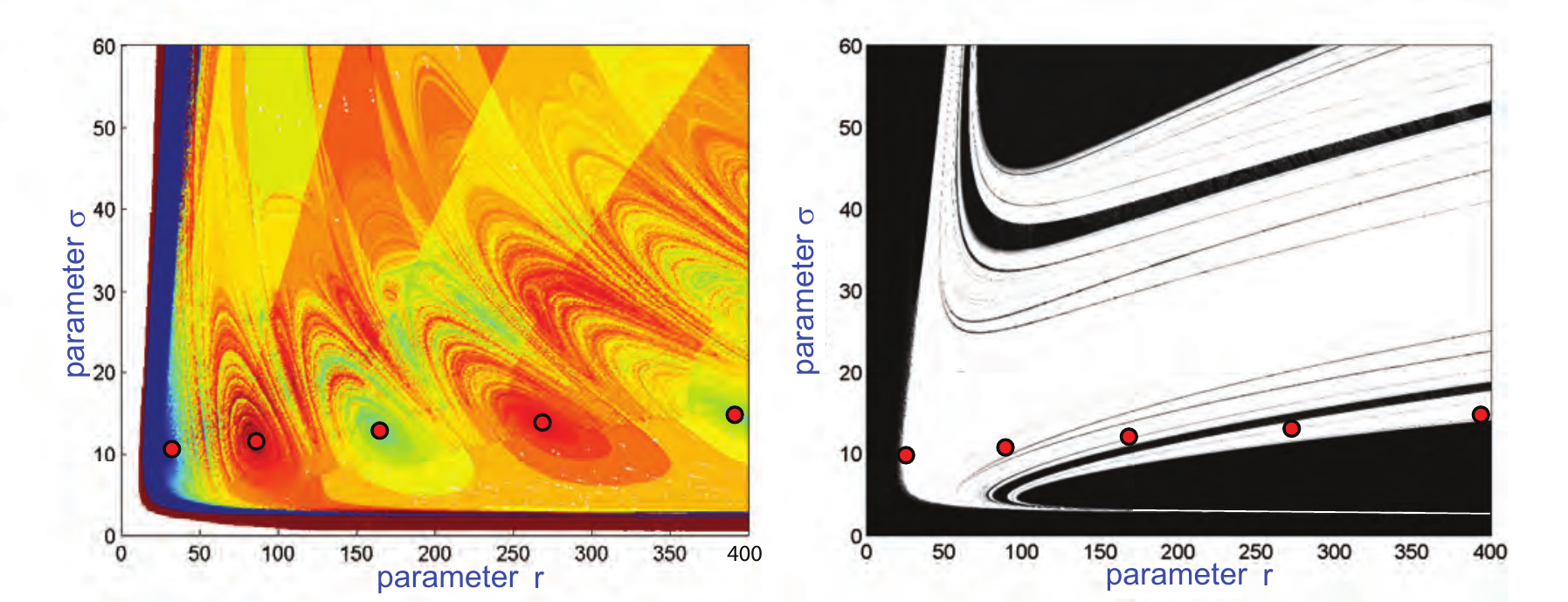}
\caption{\label{fig10} (A)  Kneading-based scan revealing a fractal structure and
a chain of spiral vortices centered at T-points alternating with saddles in the  extended $(r,\sigma)$-region of the Lorenz equation.
 (B) $L_{\rm max}$-based sweeping of the Lorenz equation singes out stability windows (dark) corresponding to steady state and emergent periodic attractors with $L_{\rm max} \le 0$ within the chaotic region (white) associated with $L_{\rm max}>0$.}
 \end{center}
\end{figure*}

Two panels in Fig.~\ref{fig10} represent the expanded scans of the of Lorenz equation:  panel~(A) is the  kneading invariant mapping on
the grid of $[1000 \times 1000]$ points and panel~(B) shows $L_{max}$-based sweeping of the $(r,\sigma)$-parameter plane. While panel~A
reveals a chain of large-scale spiral hubs around T-points, panel~B reveals none but stability windows (dark). We would like to point out
that the stability windows can also be detected in the kneading scan in panel~A showing the border of the solid color (dark yellow) island
stretched horizontally at small values of the parameter $\sigma$.

\section{The Shimizu-Morioka model}
\label{sec:5}

Next we perform the kneading-based bi-parametric scanning of another classical three-dimensional system
called the Shimizu-Morioka model \cite{SM80,ALS86,ALS89,ALS91,ASHIL93}:
\begin{equation}
\dot{x} = y, \quad \dot{y} = x -\lambda y -xz, \quad \dot{z} = - \alpha z+ x^2; \label{sm}
\end{equation}
here, $\alpha$ and $\beta$ are positive bifurcation parameters. Like the Lorenz equation, this $\mathbb{Z}_2$-symmetric model has 3 equilibrium
states: a saddle of the (2,1)-topological type at the origin, and two symmetric stable-foci or saddle-foci of the (1,2)-topological type.

\begin{figure*}[ht!]
 \begin{center}
\includegraphics[width=.6\textwidth]{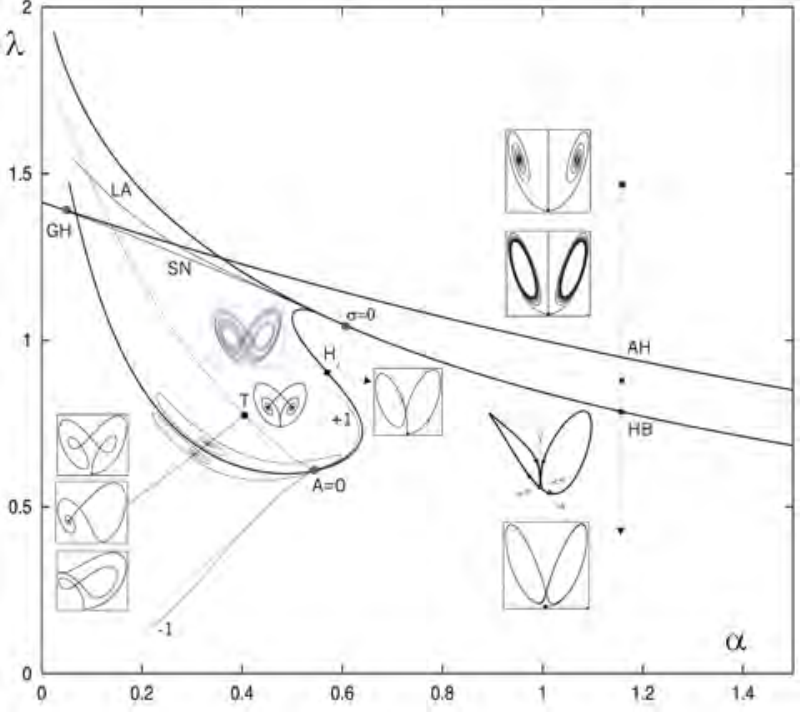}
\caption{\label{sm1} A partial ($\alpha,\,\lambda$)-diagram of the Shimizu-Morioka model depicting initial bifurcation curves
and corresponding insets for the separatrix behaviors. Legend: $AH$ stands for a supercritical Andronov-Hopf
bifurcation, $HB$ stands for the homoclinic butterfly made of two separatrix loops; $SN$ stands  for a saddle-node bifurcation
of periodic orbits; it connects the codimension-two points, the resonant saddle $\sigma=0$ on $HB$ and the Bautin bifurcation at $GH$.
$LA$ stands for the Lorenz attractor formation;  $A=0$ stands for an orbit-flip bifurcation for the double-loop homoclinics on $H_2$.
The dashed line separates, with good precision, the Lorenz attractor region from the region of a quasi-attractor (below). Vertical Pathway showing the gluing bifurcation on $HB$. Courtesy of \cite{SSTC01}.}
 \end{center}
 \end{figure*}

This model was originally introduced to examine a pitch-fork bifurcation of the stable figure-8 periodic orbit that gives rise
to multiple cascades of period doubling bifurcations in the Lorenz equation at large values of the Reynolds number $r$.
It was proved in \cite{SST93} that the Eqs.~(\ref{sm}) are a universal normal form for several
codimension-three bifurcations of equilibria and periodic  orbits on $\mathbb{Z}_2$-central  manifolds.
The model turned out to be very rich dynamically: it exhibits various interesting global bifurcations
\cite{ALS86,ALS91,ASHIL93} including T-points for heteroclinic connections.

While the model inherits all basic properties of the Lorenz equation, in addition, and of special interest, are two homoclinic bifurcations of
codimension-two: resonant saddle with the zero saddle value or the saddle index $\nu=1$, and the orbit-flip bifurcation where
the separatrix value $A$ in Eq.~(\ref{map}) vanishes. Recall that the sign of the separatrix value determines whether the homoclinic
loop, here double-pulsed, is oriented or flipped. These codimension-two points  globally organize the structure
of the compact ($\alpha,\,\lambda$)-parameter region of the Shimizu-Morioka model, including
structural transformations of the Lorenz attractor in the model, including
the emergence of stability islands -- lacunae inside the attractor.

\begin{figure}[p]
 \begin{center}
\includegraphics[width=0.95\textwidth]{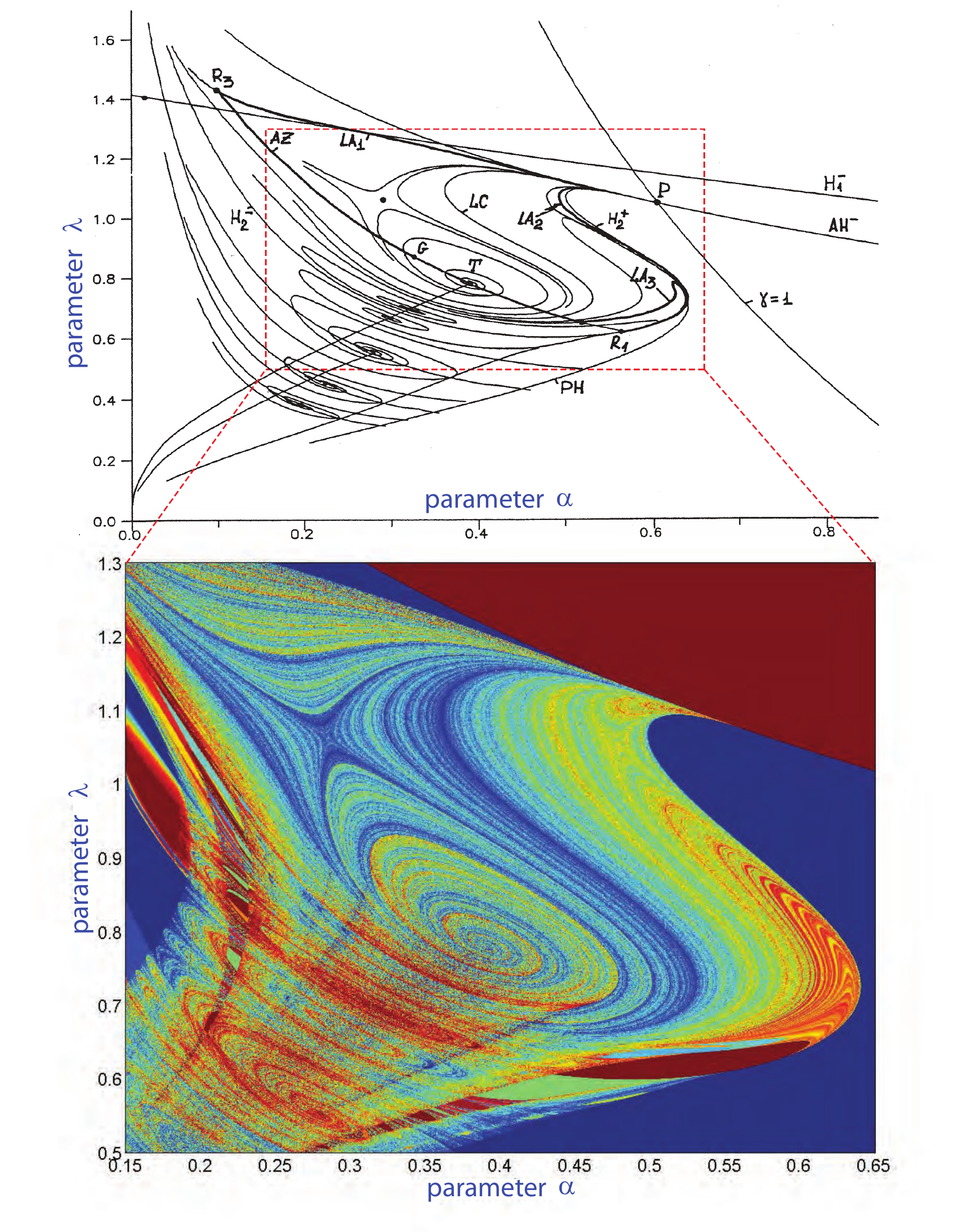}
\caption{\label{fig5l} Detailed $(\alpha,\lambda)$-parameter plane of the Shimizu-Morioka model obtained by the
parameter continuation method (courtesy of \cite{SST93}) and by the bi-parametric scan based on the kneading invariants.
The scan revealing multiple T-points and saddles that  globally organize
complex chaotic dynamics of the model. Solid-color regions associated with constant values of the kneading invariant correspond to simple
dynamics dominated by stable equilibria (brown) or stable periodic orbits (blue). The border between
the brown and blue regions corresponds to the bifurcation curve of the homoclinic butterfly. The codimension two
point, $\sigma=0$, gives rise to loci of bifurcation curves including $LA_1$ below which the Lorenz attractor exists.
Bifurcation loci of the other codimension two point, $A=0$ (orange zone)  giving rise to subsidiary orbit-flip bifurcations on turns of spirals
around T-points, are separated by saddles (two large scale ones) in the parameter plane. }
 \end{center}
\end{figure}

Figure~\ref{fig5l} represents a partial ($\alpha,\,\lambda$)-diagram of the Shimizu-Morioka model and depicts
primary bifurcation curves and the corresponding phase portraits of the separatrix behaviors. Among those are the Andronov-Hopf
bifurcation curve, $AH$, above which the equilibrium states, $O_{1,2}$ are stable, and saddle-foci below. This bifurcation
is primarily supercritical, but becomes subcritical at smaller values of  the parameter $\alpha$.
The bifurcation curve, $HB$, corresponds to the formation of the homoclinic butterfly, or figure-8 at larger
values of $\alpha$. The transition between the branches of the curve is no bifurcation like
the inclination-switch because the saddle value, $\sigma$ is negative here. The pathway at $\alpha=1.15$ demonstrates the evolution of the simple
Morse-Smale dynamics of the model from the stable equilibrium states to a stable symmetric periodic orbit. This orbit emerges through
a gluing bifurcation after that two stable periodic orbits existing below the supercritical Andronov-Hopf curve $AH$ form the
homoclinic butterfly with $\sigma<0$.
The saddle value becomes positive to left from the codimension-two point, labeled by $\sigma=0$ on the bifurcation curve $HB$.
The left segment of the curve $HB$ is similar to the bifurcation curve, $l_1$, of the Lorenz equation (Fig.~\ref{fig2l}); namely,
the homoclinic butterfly with $\sigma>0$ causes the homoclinic explosion in the phase space of the Shimizu-Morioka model.
The curve $LA$ ($LA_1$ in Fig. \ref{fig5l}) being an analog of the curve $l_2$ in the bifurcation diagram in Fig.~\ref{fig2l} is the upper
boundary of the existence region of the Lorenz attractor is the parameter plane of this model. Below $LA$ the separatrices of the saddle
no longer converge to stable periodic orbits but fill in the strange attractor, like in Fig.~\ref{fig2ln}.
The bifurcation unfolding of the homoclinic resonant saddle includes various bifurcation curves: among them in the figure we depict, in addition to $LA$, the curve, $SN$, corresponding to saddle-node  of merging stable (through the supercritical Andronov-Hopf bifurcation) and saddle
(through the homoclinic bifurcation on $HB$) periodic orbits. The curve labeled by $H_2$ corresponds to a pair of
the double-pulsed homoclinic loops. Continued further away from the codimension-two point, $\sigma=0$, the curve $H_2$ frames the chaotic region
in the parameter plane. The point, $A=0$, on this curve is a codimension-two orbit-flip homoclinic bifurcation: to the left of it, the loops become
flipped like the median of a M\"obius band. The dashed line, $AZ$ originating from the point $A=0$ is alike the curve $l_k$ in Fig.~\ref{fig4l}
for the Lorenz equation.  This curve is the second boundary of the existence region (above) of the Lorenz attractor in the Shimizu-Morioka model,
that separates wild chaos with homoclinic tangencies (below). This like passes through the primary T-point in the $(\alpha, \lambda)$-parameter
plane. Figure~\ref{fig5l} depicts a few more bifurcation curves originating from the point, $A=0$: two four-pulses
homoclinic curve terminating at subsidiary T-points and the curve labeled by ``$-1$'' corresponding to a period doubling bifurcation.
An intersection of the dashed line with a homoclinic bifurcation curve (see Fig.~\ref{fig5l})
corresponds to another orbit-flip bifurcation and so forth.

\begin{figure}[ht!]
 \begin{center}
\includegraphics[width=.45\textwidth]{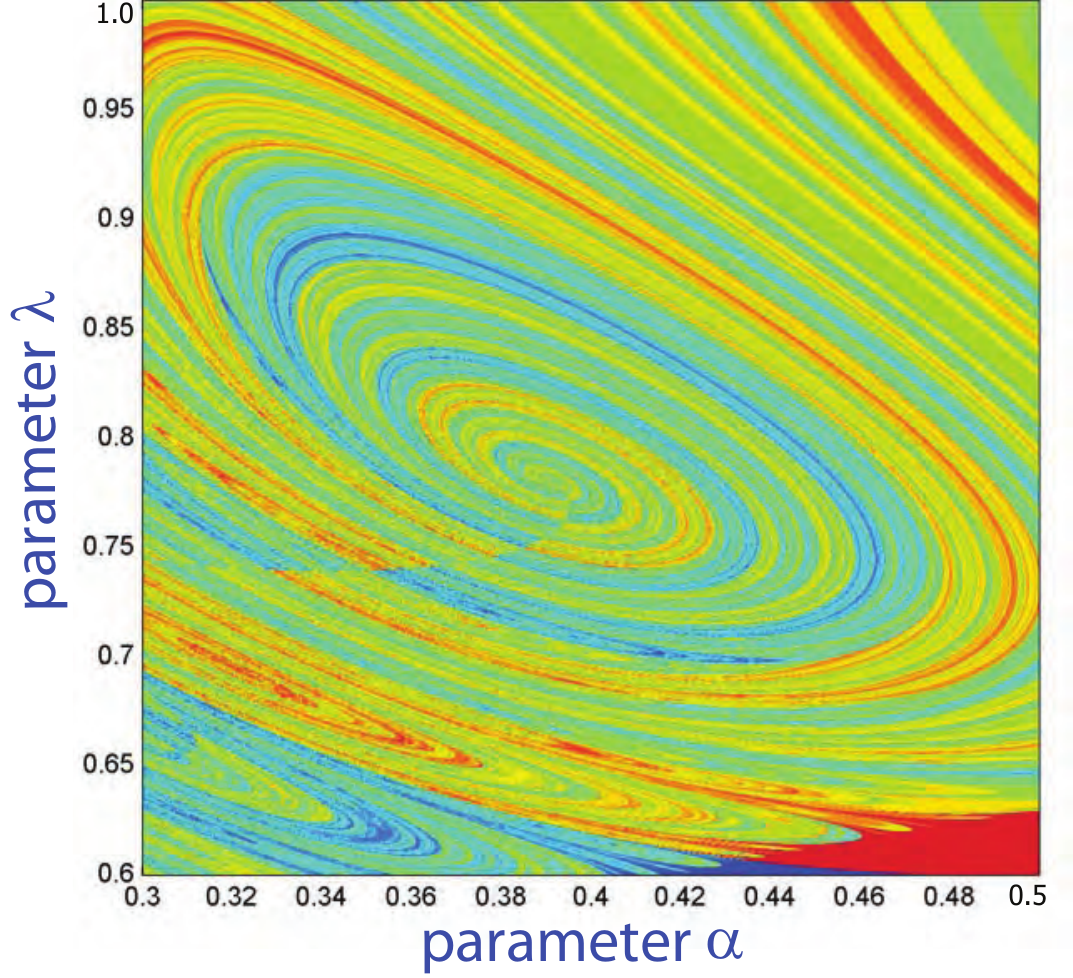}
\caption{\label{fig6l} Zoom of the $(\alpha,\lambda)$-parametric mapping in Fig.~\ref{fig5l}(B) near the primary
T-point  revealing self-similar structures embedding smaller-scale spirals around secondary T-points in the
Shimizu-Morioka model.}
\label{zoom}
\end{center}
\end{figure}

Indeed the skeleton of the bifurcation set of the Shimizu-Morioka is more complex. The detailed bifurcation diagram
is shown in the top panel of Fig.~\ref{fig5l}. It reveals several T-points, the pitch-fork bifurcation curve, $PH$, among other
bifurcation curves for various homoclinic and heteroclinic connections.  The detailed description of the bifurcation structure
of the Shimizu-Morioka model is out of scope of this paper. The curious reader can find a wealth of information on bifurcations
of the Lorenz attractor in the original papers \cite{ALS86,ALS89,ASHIL93,SST93}. We point out that those bifurcation
curves were continued in the $(\alpha,\lambda)$-parameter plane  of the model using A. Shilnikov's home-made software also based on
the symbolic kneading toolbox.

The bottom panel of Fig.~\ref{fig5l} is a {\em de-facto} proof of the new kneading invariant mapping technique.
The panel represents the bi-parametric scan in color of the dynamics of the Shimizu-Morioka model that is based on the evaluation of
the first 50 kneadings of the separatrix of the saddle on the grid of $1000\times 1000$ points in the $(\alpha,\lambda)$-parameter region.
Getting the mapping took about one hour on a high-end workstation without any parallelization efforts.
The color scan reveals a plethora of large-scale T-points, as well as nearby smaller ones (Fig.~\ref{fig6l}) invisible in the given
parameter range, as well as the saddles separating spiral structures.

The solid-color zones in the mapping correspond to simple Morse-Smale dynamics in the model. These trivial dynamics are due to either
the separatrix converging to the stable focus $O_1$ ($O_2$) and emergent periodic orbit with the same kneading invariant
(brown region), or to the symmetric and asymmetric stable figure-8 periodic orbits (dark blue region).
The borderlines between the simple and complex dynamics in the Shimizu-Morioka model are clearly demarcated. On the top it is
the curve, $LA_1$, (see the top panel of Fig.~\ref{fig5l}). The transition from the stable 8-shaped periodic orbits to the Lorenz
attractor (through the boundary, $LA_{2}$) is similar though more complicated as it involves a pitch-fork bifurcation and
bifurcations of double-pulsed homoclinics, see \cite{ASHIL93,SST93} for details.

\begin{figure*}[ht!]
 \begin{center}
\includegraphics[width=.5\textwidth]{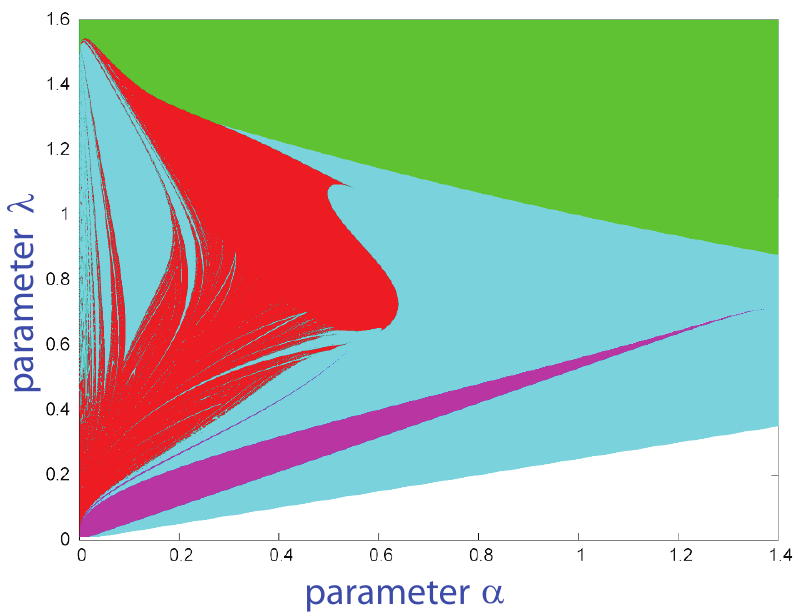}~~\includegraphics[width=.5\textwidth]{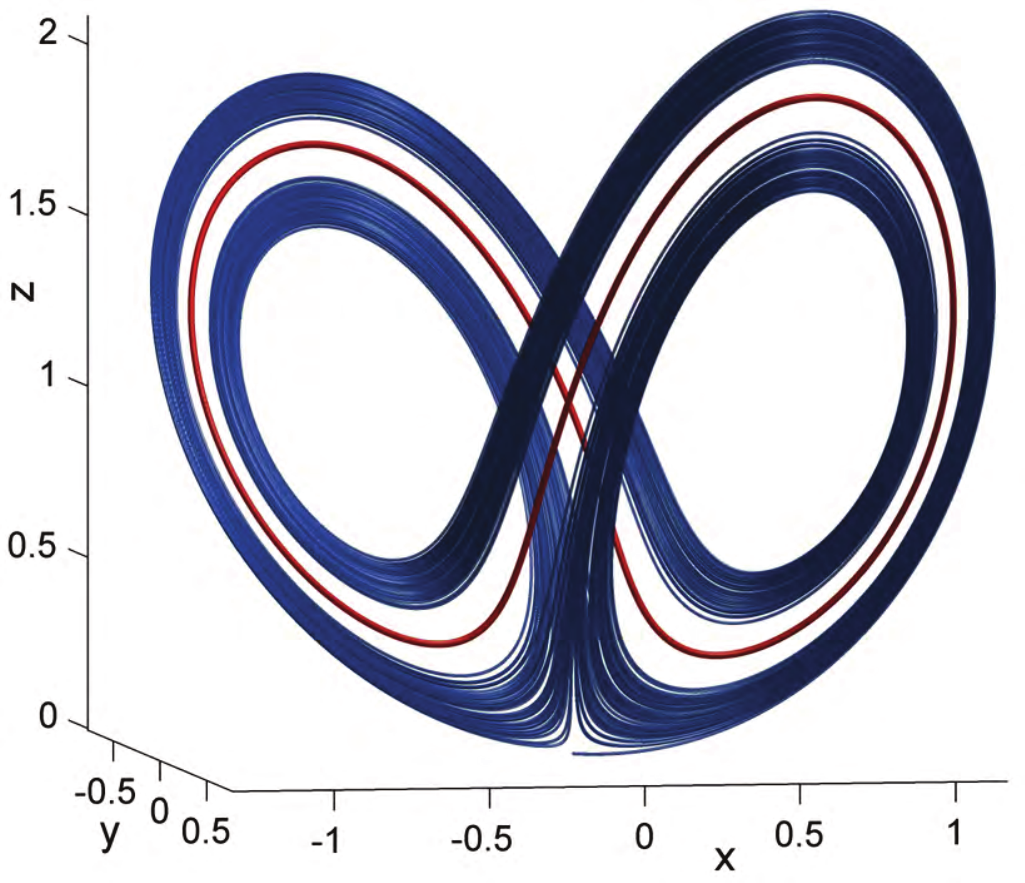}
\caption{\label{fig7l} (A) Attractors of the Shimizu-Morioka model being differentiated by the sign of the largest Lyapunov exponent, $L_{max}$.
Color legend for the attractors of the model: green--stable equilibrium states, $L_{max}<0$; blue --
stable periodic orbits with a nodal normal behavior, $L_{max}=0$; magenta -- a periodic orbit with a focal normal
behavior; red -- chaotic attractors with $L_{max}>0$, with identified lacunae.
Courtesy of \cite{GOST05}. (B)  Lorenz attractor (blue) with a lacuna containing a symmetric figure-8 periodic orbit (dark red).}
 \label{simo}
 \end{center}
 \end{figure*}

One can clearly see the evident resemblance between both diagrams found using the bifurcationaly exact numerical methods and
by scanning the dynamics of the model using the proposed kneading invariant technique. The latter reveals
a richer structure providing finer details. The structure can be enhanced further by examining longer tails of the kneading
sequences. This allows for the detection of smaller-scale spiral structures within scrolls of the primary T-vortices, as
predicted by the theory \cite{BYK93}. Figure~\ref{zoom} shows a magnification of the scan of the Shimizu-Morioka model near the
primary T-point that depicts several other small-scale T-points.

Finally, we compare the new kneading scanning apparatus with the customary bi-parametric sweeping (shown in Fig.~\ref{simo}) of the Shimizu-Morioka model
that is based on the evaluation of the Lyapunov exponent spectrum computed over a finite time interval \cite{GOST05}. Likewise the case of the
Lorenz model, the sweeping based on the Lyapunov exponents shows no sign of spiral or saddle structures inside the region of deterministic chaos.
The regions of the solid colors are associated with the sign of the largest Lyapunov exponent, $L_{\rm max}$: negative $L_{\rm max}$ values correspond
to steady state attractors in the green region; $L_{\rm max}=0$ corresponds to periodic attractors in the blue region; and $L_{\rm max}>0$  is
associated with chaotic dynamics in the model in the region. Note blue islands in the red-colored region that correspond to
stability windows in chaos-land. In such windows the Lorenz attractor has an emergent lacuna containing, initially, a single symmetric saddle periodic
orbit. The orbit then undergoes a pitch-fork bifurcation that makes it stable. The basin of the stable orbit, which is first bounded by the 2D stable manifold of two asymmetric saddle periodic orbits, increases so that the stable orbit starts to dominate over chaotic dynamics in the corresponding stability
window. These bifurcations underlie the transition from simple dynamics (blue region) due to the symmetric stable periodic orbit
to chaos through the curve, $H_2$, as the parameter $\alpha$ in decreased.

\section{6D optically pumped laser model}
\label{sec:6}

The coexistence of multiple T-points and accompanying fractal structures in the parameter plane is a signature for
systems with the Lorenz attractor. A question though remains whether the new computational technique will work for systems of dimensions higher
than three. In fact, to apply the technique to a generic Lorenz-like system, only wave forms of a symmetric variable progressing in time,
that consistently starts from the same initial condition near the saddle is required. Next is an example from nonlinear optics  --
a 6D model of the optically pumped, far infrared red three-level molecular laser \cite{Moloney1989,FMH91} given by
\begin{equation}
\begin{array}{lll}
\dot{\beta} &=& -\sigma \beta + g p_{23}, \\
\dot{p}_{21} &=& -p_{21}-\beta p_{31}+\alpha D_{21}, \\
\dot{p}_{23} &=& -p_{23}+\beta D_{23} -\alpha p_{31}, \\
\dot{p}_{31} &=& -p_{31}+\beta p_{21} + \alpha p_{23}, \\
\dot{D}_{21} &=& -b(D_{21}-D_{21}^0) - 4 \alpha p_{21} -2 \beta p_{23}, \\
\dot{D}_{23} &=& -b(D_{23}-D_{23}^0) - 2 \alpha p_{21} -4 \beta p_{23}.
\end{array}\label{laseH}
\end{equation}
Here, $\alpha$ and $b$ are the Rabi flopping quantities representing the electric field amplitudes at pump and emission frequencies. The parameter $\alpha$ is a natural bifurcation parameter as it is easily varied experimentally. The second bifurcation parameter,  $b$, can be varied to some degree at the laboratory by the addition of a buffer gas. This system presents, like the Lorenz equations, a symmetry
$(\beta, p_{21}, p_{23}, p_{31}, D_{21}, D_{23}) \leftrightarrow  (-\beta, p_{21}, -p_{23}, -p_{31}, D_{21}, D_{23})$. The laser model has either a single central equilibrium state, $O$ (with $\beta = 0$),  or additionally, through a pitch-fork bifurcation, a pair of symmetric equilibrium states, $O_{1,2}$ (with  $\beta \ge 0$);  the stability of the equilibria depends on the parameter values.

\begin{figure}[!htb]
 \begin{center}
\includegraphics[width=.5\textwidth]{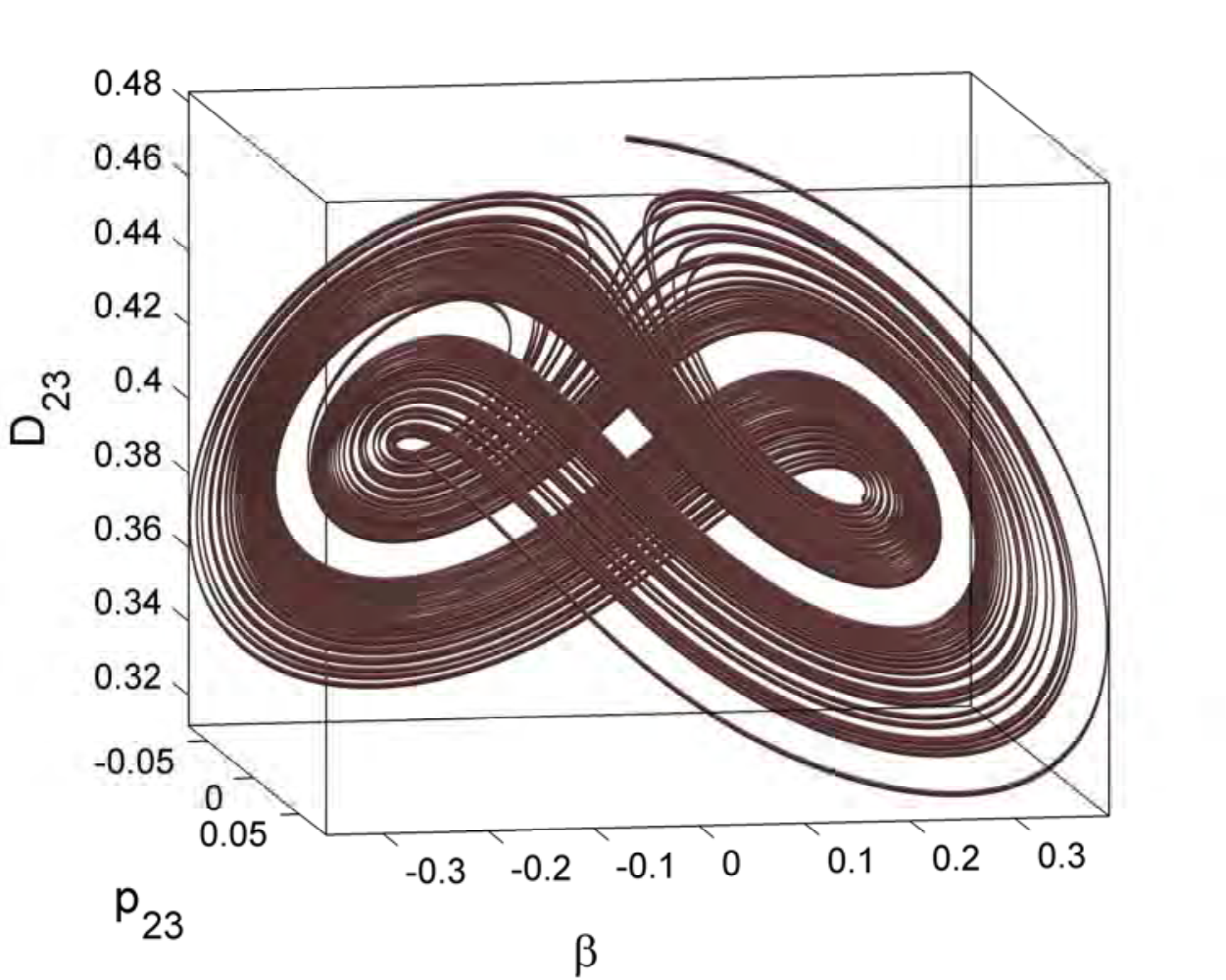}~\includegraphics[width=.45\textwidth]{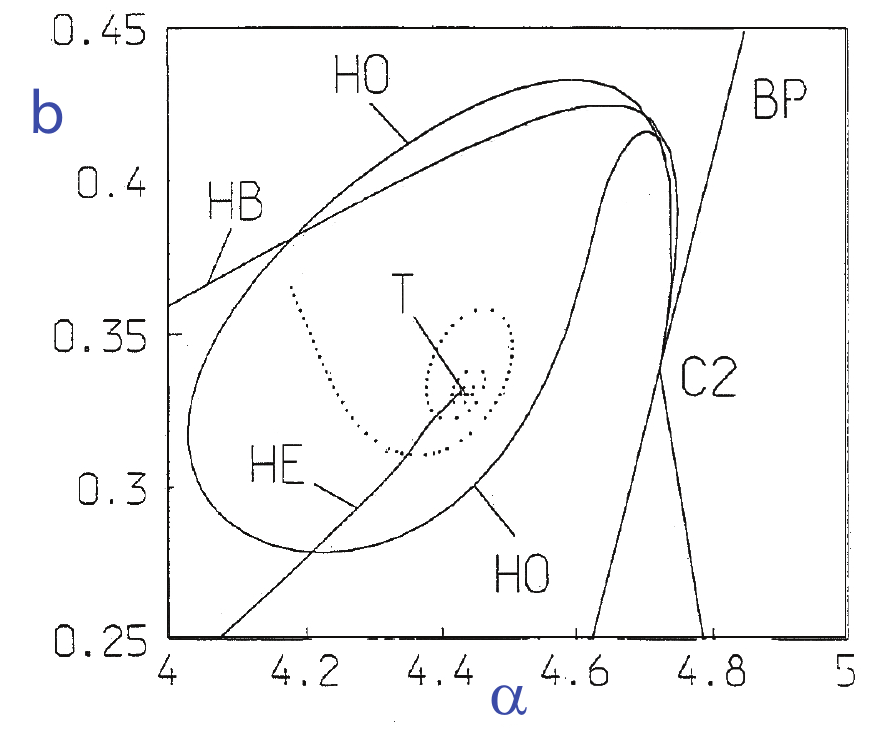}
\caption{\label{laser1} (A) Lorenz attractor with a lacuna in the laser model at $a=1.14$, $b=0.2$, $q=50$ and $\sigma=10$.
 (B) $(\alpha, b)$-bifurcation diagram of the model for $g=52$ and $\sigma=1.5$. $BP$ and HB here denote the pitch-fork and Andronov-Hopf bifurcations, respectively. $HO$ and $HE$ denote the branches of the primary homoclinic (of the saddle) and heteroclinic orbits (of the saddle-foci). $C2$ is the codimension two Khorozov-Taken point for the equilibrium state with double zero eigenvalues, and $T$ is the primary terminal point. The spiraling curve
connects the T-point with the homoclinic resonant saddle on $HO$, near which separatrix loops are double pulsed ones. Courtesy of \cite{FMH91}.
}
 \end{center}
\end{figure}

Optically pumped, far infrared lasers are known to demonstrate a variety of nonlinear dynamic behaviors, including Lorenz-like chaos \cite{laser95}. An acclaimed
example of  the modeling studies of chaos in nonlinear optics is the two level laser Haken model \cite{Haken75} to which the Lorenz equation can be reduced. A validity that three level laser models would inherently persist the Lorenz dynamics  were widely questioned back then.
It was comprehensively demonstrated in  \cite{FMH91} in 1991 that this plausible laser model possesses a plethora of dynamical and structural
features of the Lorenz-like systems, including the Lorenz attractor {\it per se} (with lacunae as well), similar Andronov-Hopf, $\mathbb{Z}_2$ pitchfork, various homoclinic and heteroclinic bifurcations including codimension-two ones  T-points, see Fig.~\ref{laser1}. Similar structures were
also discovered in another nonlinear optics model for a laser with a saturable absorber which can be reduced to the Shimizu-Miorioka model
near a steady state solution with triple zero exponents \cite{VV93}

\begin{figure}[!htb]
 \begin{center}
\includegraphics[width=1.\textwidth]{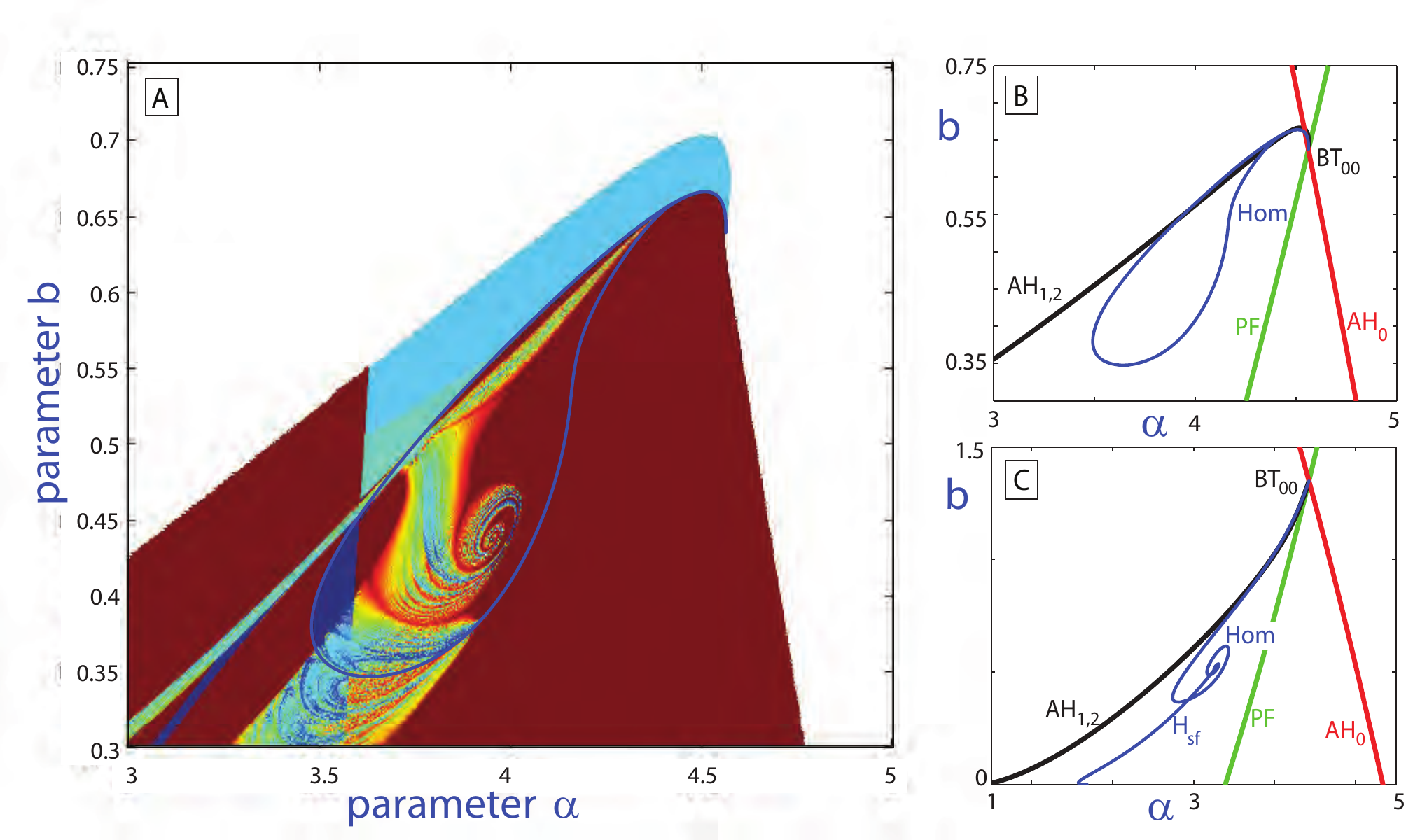}
\caption{\label{laser2} (A) Bi-parametric scan of the laser model featuring the T-points and saddles
typical for the Lorenz-like systems, mapping the dynamics of the 6D optically pumped far-infrared-red laser model onto
the (electric-field-amplitude, omission-frequency)-diagram  at $g=50$ and $\sigma = 1.5$. Solid-color windows and fractal
regions correspond to trivial and chaotic dynamics generated by the laser model. (B) Partial bifurcation diagram though the parameter
continuation showing the curves for pitch-fork ($PF$) and Andronov-Hopf ($AH_0$) bifurcations for the equilibrium state, $O$, and
another similar supercritical one for  $O_{1,2}$. The homoclinic curve, $Hom$ begins from the codimension-two point, $BT$ for the Khorozov-Takens
bifurcation and ends up at the resonant saddle point. (B) Elevating $\sigma =2$ makes the $Hom$ turned by a saddle point in the parameter plane
and terminate at the primary T-point.}
 \end{center}
\end{figure}

The laser model~(\ref{laseH}) is quite rich in bifurcations; the list also includes a super-critical Andronov-Hopf bifurcation of the  central equilibrium state that gives rise to a stable figure-8 shaped periodic orbit (in proper projections)
for the parameter values to the left of the bifurcation curve, $AH_0$, in the
bifurcation diagrams shown in panels B and C of Fig.~\ref{laser2}. Observe from the diagram that the curve $AH_0$  originates from the point labeled, $BT$. This point corresponds to a codimension-two Khorozov-Takens bifurcation of an equilibrium state with two zero Lyapunov exponents. The bifurcation
is an extension of the Bogdadov-Takens bifurcation on a symmetric central manifold. The unfolding of this bifurcation includes two more curves:
 $AH_{1,2}$ standing for the Andronov-Hopf bifurcation for the secondary equilibrium states, $O_{1,2}$; and $Hom$ standing for a homoclinic connection made of two symmetric separatrix loops bi-asymptotic to the saddle, $O$. The continuation of the curve, $Hom$, away from $BT$ reveals rather peculiar  details
 that  substantially organize the bifurcation diagram of this laser model. Near $BT$ the curve, $Hom$, corresponds to a homoclinic figure-8  of the saddle with a negative saddle value,  on the $\mathbb{Z}_2$-symmetric 2D central manifold in the 6D phase space of the model.  Recall that the figure-8 homoclinic connection stands for the case where the 1D unstable separatrices come back to the saddle, $O$ from the symmetrically opposite directions along the eigenvector corresponding to the leading stable exponent at the equilibrium state.  This bifurcation gluing two stable periodic orbits emerging from $O_{1,2}$ through the supercritical Andronov-Hopf bifurcation gives rise to the stable symmetric figure-8 periodic orbit existing nearby $BT$. As the curve,  $Hom$, is continued further away from  $BT$, the stable leading direction at the saddle, $O$, changes: it becomes the
invariant $\beta$-axis (like the $z$-axis in the Shimizu-Morioka model) so that the separatrix loops start returning tangent to each other and hence forming  the homoclinic butterfly. Nevertheless, this is a gluing bifurcation, not a codimension-two bifurcation of the change of the leading direction (inclination switch) as the saddle value remains negative on this branch of the curve, $Hom$. The saddle value vanishes, making the saddle resonant at the codimension-two point,  and becomes positive further down on $Hom$. As the curve is continued, the homoclinic butterfly undergoes another codimension-two orbit-flip bifurcation so that the separatrices loops of the saddle, $O$ become non-orientated. As the result, further down the curve, each loop gains an extra turn around the incoming separatrix of the opposite saddle-focus, i.e. becomes a double-pulsed one with the  $\{1,-1,0\}$ kneading. Depending on the parameter cut $\sigma$ there are two scenarios for termination of the curve, $Hom$, in the $(\alpha, b)$ diagram (compare the bifurcation diagrams in panels B and C of Fig.~\ref{laser2}): first, when $\sigma=1.5$ it terminates at the codimension-two T-point corresponding to the heteroclinic connection between all saddle equilibria, $O$ and $O_{1,2}$ as shown in panel B. The second scenario for is less predictable at $\sigma=2$: the branch, $Hom$, of double-pulsed separatrix loops ends up at the codimension-two point of the resonant saddle with the zero saddle-value (panel C).
The answer to the question what makes the curve change its destination is a saddle point in the parameter diagram that the kneading scan  reveals in Fig.~\ref{fig5l}. By varying the $\sigma$-parameter cut in the 3D parameter space, this bifurcation curve is destined by the saddle to finish at either terminal point, see details in \cite{SST93}. In the case where  it spirals onto the T-point, there is another bifurcation curve  corresponding to the same $\{1,-1,0\}$ kneading  that connects the codimension-two orbit-switch point and the point corresponding to the resonant saddle located on the curve $Hom$.

The panel A in Fig.~\ref{laser2} represents the kneading scans of the dynamics of the laser model which is mapped onto the $(\alpha, \beta)$-parameter plane with $g=50$ and $\sigma = 1.5$. The scan is done using the same 50 kneading entries. It has the regions of chaotic dynamics clearly demarcated from the solid color windows of persistent kneadings corresponding to trivial attractors such as stable equilibria and periodic orbits. The region of chaos has a vivid fractal spiral structure  featuring a chain of T-points. Observe also a thin chaotic layer bounded away from the curve $Hom$ by a curve of double-pulsed homoclinics with the kneading $\{1,-1, 0\}$ connecting the codimension-two points:  the resonant saddle and the orbit-flip both on $Hom$. One feature of these points is the occurrence of the Lorenz attractor with one or more lacunae \cite{ABS83,ALS86,ASHIL93,SST93}. Such a strange attractor with a single lacuna containing a figure-8 periodic orbit in the phase space of the given laser model is shown in panel~A of Fig.\ref{laser1}.

\section{Conclusions}

\begin{figure}[!htb]
 \begin{center}
\includegraphics[width=0.85\textwidth]{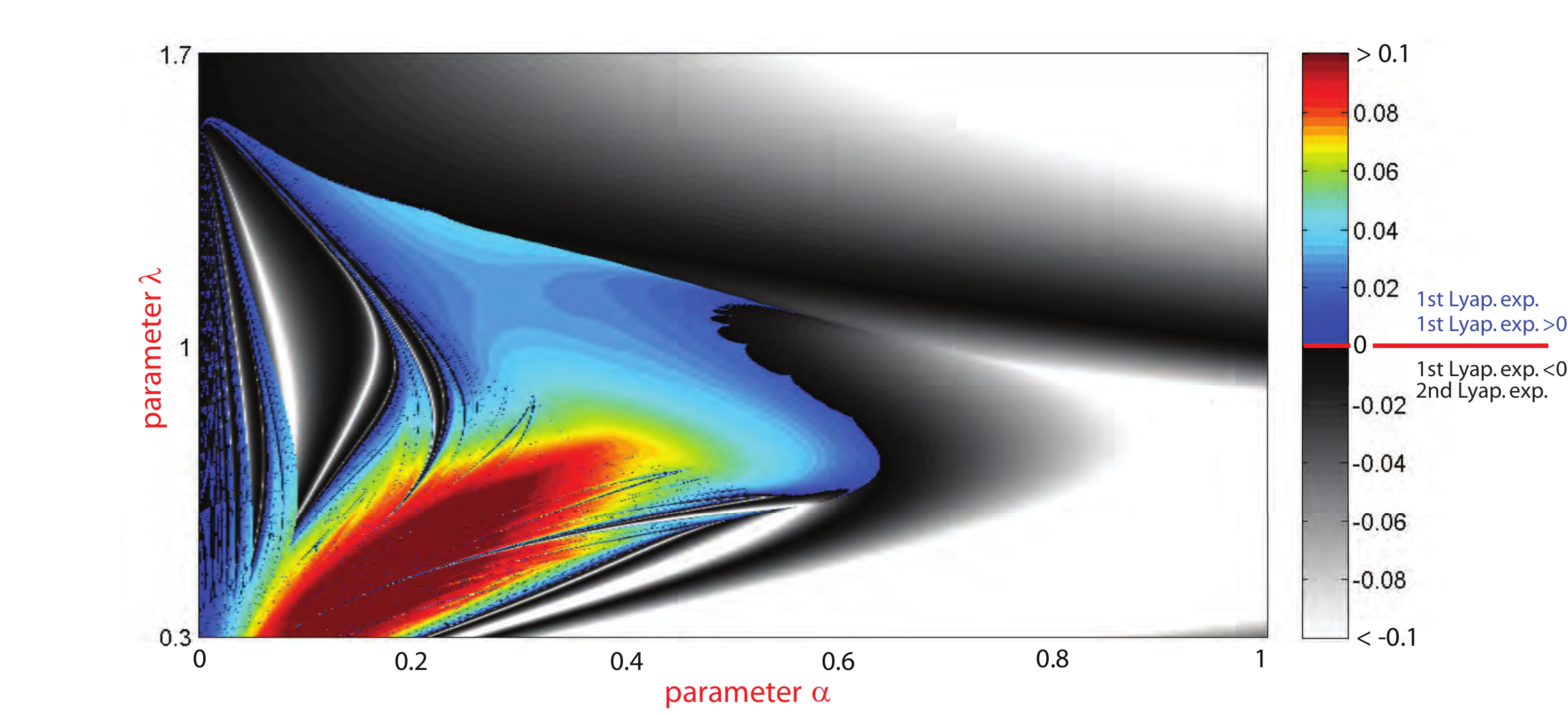}
\caption{\label{sm3} Fine scan based on the Lyapunov exponents indicating the presence of the saddle (in the
Lorenz attractor region shown in cold blue) and spiral structures (in the reddish regions with larger Lyapunov exponents for
wild chaos due to saddle-foci) in the ($\alpha, \lambda$)-parameter space of the Shimizu-Morioka model.}
 \end{center}
\end{figure}

We have demonstrated the new proposed computational toolkit for thorough  explorations of chaotic dynamics in three exemplary models with the Lorenz attractor. The algorithmically easy though powerful toolkit in based on the scanning technique that maps the dynamics of the system onto the bi-parametric plane. The core of the approach is the evaluation of the kneading invariants for regularly or chaotically varying flip-flop patterns of a single trajectory -- the separatrix of the saddle singularity in the system. In the theory, the approach allows two systems with the structurally unstable Lorenz attractors to be conjugated with the mean of a single number -- the kneading invariant. By using ready-for-use tools in Matlab, we could have the parameter plane of the model in question be foliated by the level curves of distinct colors corresponding to distinct values of the kneading invariants. The kneading scans revel unambiguously the key features in the Lorenz-like systems such as a plethora of underlying spiral structures around T-points, separating saddles in intrinsically fractal regions corresponding to complex chaotic dynamics.  We point out that no other techniques, including approaches based on the Lyapunov exponents, can reveal the discovered parametric chaos with such stunning clarity and beauty. Figure~\ref{sm3} for the Shimizu-Morioka model shows that a fine scan based on the finite-time Lyapunov exponents is able to indicate some presence of spiral and saddle structures and differentiate between the chaotic dynamics due to the Lorenz attractor and those due to additional degrees of instability brought in by saddle-foci.

The kneading based methods should be beneficial for detailed studies of other systems admitting reasonable symbolic descriptions.
It bears an educational aspect as well: the kneading based scanning can be used for in-class presentation to reveal
the fascinating complexity of low-order models in the cross-disciplinary field of nonlinear dynamics. The bi-parametric mapping
technique can be easily adopted for parallel multicore GPU platforms allowing for ultra-fast simulations of models in questions.
Additional implementation  of high-precision computations of long transients shall thoroughly reveal multi-layer complexity of self-similar
fractal patterns near T-point vortices. In future research we intend to enhance and refine the toolkits for exploration of other symmetric and
asymmetric \cite{ALSLP91} systems of differential and difference equations, like 3D Poincar\'e mappings \cite{SST93,GOST05}, including 4D models
with saddle-foci, that require two and more kneading invariants for the comprehensive symbolic description.

\section{Acknowledgments}

This work is supported by the Spanish Research project MTM2009-10767 (to R.B.), and by NSF grant DMS-1009591, RFFI Grant No.~08-01-00083,
GSU Brain \& Behaviors program, and MESRF project 14.740.11.0919 (to A.S) as well as
by the Grant 11.G34.31.0039 of the Government of the Russian Federation
for state support of scientific research conducted under supervision of leading scientists in
Russian educational institutions of higher professional education  (to L.P.S).
We thank Dima Turaev for stimulating discussions,  Yuri Kuznetsov for sharing the data used in Fig.~\ref{fig3l}.
We thank Aaron Kelley and Jeremy Wojcik for careful proofreading the manuscript.


\end{document}